\documentclass[]{bytedance_seed}

\usepackage[toc,page,header]{appendix}

\usepackage{minitoc}

\usepackage{amsmath,amsfonts,bm}
\usepackage{booktabs}
\usepackage{graphicx}
\usepackage{multirow}
\usepackage{makecell}
\usepackage{enumitem}
\usepackage{subcaption}
\usepackage{diagbox}
\usepackage{pifont}
\usepackage{colortbl}
\usepackage{threeparttable}
\usepackage{tabularx}
\usepackage{float}

\usepackage[most]{tcolorbox}
\usepackage{framed}

\definecolor{lightgreen}{rgb}{0.88, 1, 0.88}
\definecolor{lightred}{rgb}{1, 0.88, 0.88}
\definecolor{lightyellow}{rgb}{1, 1, 0.88}
\definecolor{lightblue}{rgb}{0.88, 0.95, 1}
\definecolor{lightorange}{rgb}{1, 0.93, 0.88}

\definecolor{yushunred}{rgb}{0.82, 0.18, 0.14}
\definecolor{yushunblue}{rgb}{0.16, 0.33, 0.68}
\definecolor{yushungreen}{rgb}{0.31, 0.65, 0.31}

\usepackage{CJKutf8}  

\usepackage[most]{tcolorbox}
\newtcolorbox{promptbox}{
    colback=gray!5,
    colframe=gray!75,
    boxrule=0.5pt,
    arc=4pt,
    left=6pt,
    right=6pt,
    top=6pt,
    bottom=6pt,
    breakable,
    fontupper=\small  
}

\title{SpeechJudge: Towards Human-Level Judgment for Speech Naturalness}

\author[1,2,*]{Xueyao Zhang}
\author[1,*]{Chaoren Wang}
\author[1]{Huan Liao}
\author[1]{Ziniu Li}
\author[1]{Yuancheng Wang}
\author[1]{Li Wang}
\author[2]{Dongya Jia}
\author[2]{Yuanzhe Chen}
\author[3]{Xiulin Li}
\author[2]{Zhuo Chen}
\author[1]{Zhizheng Wu}

\affiliation[1]{The Chinese University of Hong Kong, Shenzhen}
\affiliation[2]{ByteDance Seed}
\affiliation[3]{DataBaker Technology}

\abstract{
Aligning large generative models with human feedback is a critical challenge. In speech synthesis, this is particularly pronounced due to the lack of a large-scale human preference dataset, which hinders the development of models that truly align with human perception. To address this, we introduce \textbf{\textit{SpeechJudge}}, a comprehensive suite comprising a dataset, a benchmark, and a reward model centered on \textit{naturalness}—one of the most fundamental subjective metrics for speech synthesis. First, we present \textbf{\textit{SpeechJudge-Data}}, a large-scale human feedback corpus of 99K speech pairs. The dataset is constructed using a diverse set of advanced zero-shot text-to-speech (TTS) models across diverse speech styles and multiple languages, with human annotations for both intelligibility and naturalness preference. From this, we establish \textbf{\textit{SpeechJudge-Eval}}, a challenging benchmark for speech naturalness judgment. Our evaluation reveals that existing metrics and AudioLLMs struggle with this task; the leading model, Gemini-2.5-Flash, achieves less than 70\% agreement with human judgment, highlighting a significant gap for improvement. To bridge this gap, we develop \textbf{\textit{SpeechJudge-GRM}}, a generative reward model (GRM) based on Qwen2.5-Omni-7B. It is trained on SpeechJudge-Data via a two-stage post-training process: Supervised Fine-Tuning (SFT) with Chain-of-Thought rationales followed by Reinforcement Learning (RL) with GRPO on challenging cases. On the SpeechJudge-Eval benchmark, the proposed SpeechJudge-GRM demonstrates superior performance, achieving 77.2\% accuracy (and 79.4\% after inference-time scaling @10) compared to a classic Bradley-Terry reward model (72.7\%). Furthermore, SpeechJudge-GRM can be also employed as a reward function during the post-training of speech generation models to facilitate their alignment with human preferences.
}

\correspondence{Zhizheng Wu at \url{wuzhizheng@cuhk.edu.cn}}

\checkdata[Project Page]{\url{https://speechjudge.github.io/}}
\checkdata[Dataset, Model, and Code]{\url{https://github.com/AmphionTeam/SpeechJudge}}

\begin{document}
\maketitle



\section{Introduction}

The collection and integration of human feedback corpora for model alignment has become a critical stage in the development of modern large-scale generative models, proving indispensable in domains such as text~\citep{openai-rm,instructgpt,RLHF-anthoropic}, image~\citep{imagereward,pick-a-pic}, and video generation~\citep{visionreward,videoreward}. 

\renewcommand{\thefootnote}{\fnsymbol{footnote}}
\footnotetext[1]{Equal Contribution.}
\renewcommand{\thefootnote}{\arabic{footnote}} 

In the field of speech synthesis, \textbf{\textit{naturalness}} has long been a cornerstone subjective metric for quality assessment~\citep{naturalspeech3,seedtts,cosyvoice,qwen2.5-omni,kimi-audio}, representing one of the most general-purpose indicators of performance~\citep{tts-book-paul-taylor,tts-book-tanxu}. Prior research has explored automated speech assessment through MOS predictors~\citep{utmos,voicemos-challenge-2024} and constructed the human feedback corpora for specific attributes like the low-level acoustic quality~\citep{qualispeech}. However, a large-scale human feedback corpus centered on the holistic quality of naturalness—and a corresponding reward model trained to capture these preferences—remains a notably underexplored area. To fill this void, this paper focuses on the dimension of speech naturalness and present a three-part contribution:

\textbf{1. A Large-scale Human Feedback Dataset: \textit{SpeechJudge-Data}.}
We recruit human annotators to provide feedback on synthesized speeches, with a focus on assessing two fundamental speech aspects: \textit{intelligibility} and \textit{naturalness}. For data synthesis, we employ a diverse set of advanced, open-source zero-shot TTS models with varying architectures (such as CosyVoice2~\citep{cosyvoice2}, Ints~\citep{intp}, F5-TTS~\citep{f5tts}, and MaskGCT~\citep{maskgct}) to produce the compared speech pairs. We prepare speech references in both regular and expressive styles, construct multilingual target texts, and cover both monolingual and cross-lingual synthesis scenarios to ensure data diversity (Section~\ref{sec:data-construction}). We instruct human annotators to perform two tasks based on a speech pair (Figure~\ref{fig:system-ui}): (a) pointwise annotation of text accuracy to assess intelligibility, and (b) pairwise preference annotation to judge relative speech naturalness. This extensive effort, involving 69 labelers over two months, results in 99K annotated pairs, with each pair receiving an average of 2.49 annotations from different labelers. We believe the SpeechJudge-Data can serve as a valuable corpus for alignment research in speech synthesis (e.g., DPO alignment~\citep{dpo} or reward modeling~\citep{openai-rm,instructgpt,RLHF-anthoropic} in Section~\ref{sec:speechjudge-grm}).

\textbf{2. An Evaluation Benchmark for Speech Naturalness Judgment: \textit{SpeechJudge-Eval}.}
We design a dedicated evaluation benchmark for the task of speech naturalness judgment. The task is structured as follows: given a target text and two corresponding speech samples, a model needs to judge which one is more natural. To construct the evaluation set, we select a subset from the SpeechJudge-Data where human annotators demonstrated high inter-annotator agreement, ensuring a high-quality ground truth. We assess the naturalness judgment capabilities of a wide range of metrics and models, including Word Error Rate (WER)~\citep{whisper,funasr}, Fréchet Audio Distance (FAD)~\citep{fad}, MOS predictors~\citep{utmos,dnsmos,meta-audiobox-aesthetics}, Deepfake Detectors~\citep{aasist,audio-deepfake-verification}, and AudioLLMs~\citep{qwen2.5-omni,kimi-audio,mimoaudio,gemini2.5,gpt4o}. Our evaluations reveal that even the most capable model—specifically, Gemini-2.5-Flash~\citep{gemini2.5} in our experiments—achieved less than 70\% agreement with human preferences. This finding highlights a significant performance gap and underscores the substantial room for research and improvement in automated speech naturalness judgment.

\textbf{3. A Generative Reward Model for Speech Naturalness: \textit{SpeechJudge-GRM}.}
To develop a reward model that more effectively captures human preferences, we develop SpeechJudge-GRM, a generative reward model (GRM)~\citep{google-grm,deepseek-grm} trained on the SpeechJudge-Data. Specifically, we base our model on Qwen2.5-Omni-7B~\citep{qwen2.5-omni} and design a two-stage post-training process. During the first stage, we perform Supervised Fine-Tuning (SFT) as the ``cold start'' to improve the model's instruction-following and rationale-based reasoning capabilities. To achieve this, we leverage Gemini-2.5-Flash~\citep{gemini2.5} to generate Chain-of-Thought (CoT) data for speech naturalness judgment task. In the second stage, we focus on more challenging cases of SpeechJudge-Data, which we define as instances where Gemini-2.5-Flash fails to make the correct judgment. Treating the human-annotated labels as the verifiable reward~\citep{deepseek-r1,deepseek-grm}, we apply the GRPO-based Reinforcement Learning (RL) stage~\citep{grpo}. Our experiments demonstrate that when trained on the same data, SpeechJudge-GRM significantly outperformed the classic Bradley-Terry reward model (BTRM)~\citep{bt-rm,dpo}, achieving a higher accuracy in predicting human preferences (77.2\% for SpeechJudge-GRM vs. 72.7\% for SpeechJudge-BTRM, Table~\ref{tab:grm}). Besides, SpeechJudge-GRM also supports inference-time scaling and offers explainability through its CoT outputs. Furthermore, SpeechJudge-GRM can also be employed as an objective naturalness metric for sample selection (Figure~\ref{fig:best-of-n-by-speechjudge-grm}) or as a reward function in RL algorithms to enhance the quality of existing speech generation models (Figure~\ref{fig:post-training-tts-results}).

We will release all resources at \url{https://github.com/AmphionTeam/SpeechJudge} to facilitate future research in human-aligned speech synthesis. Audio samples are available at \url{https://speechjudge.github.io/}.

\section{Related Work}

\textbf{Human Alignment for Speech Generation}\quad
Aligning generative models with human feedback has proven crucial, a process also known as RLHF in LLMs~\citep{instructgpt,RLHF-anthoropic}. In the vision domain, many similar human preference datasets exist, such as Pick-a-Pic~\citep{pick-a-pic}, ImageReward~\citep{imagereward}, and VideoReward~\citep{videoreward}. The speech synthesis field, pioneering efforts to construct human corpora involved MOS datasets~\citep{utmos,voicemos-challenge-2024}. However, these datasets often did not use advanced TTS models for data generation, provided only the pointwise labels rather than the direct pairwise human preference, and were limited in scale. More recently, efforts have focused on building human feedback corpora centered on specific speech attributes, such as low-level acoustic quality~\citep{qualispeech}, intelligibility~\citep{intp}, or the instruction-following capabilities of spoken dialogue systems~\citep{wavreward,sagelm}. Despite this progress, a large-scale human feedback corpus built specifically around \textit{naturalness}—one of the most general-purpose and fundamental metrics for speech synthesis~\citep{tts-book-paul-taylor,tts-book-tanxu}—has remained a critical missing piece.

\textbf{AudioLLM as a Judge}\quad
Using LLMs as automated quality evaluators is a prominent topic in the textual LLM field, popularized by the ``LLM-as-a-judge'' paradigm~\citep{llm-as-a-judge}. This idea has recently been extended to the audio domain. A concurrent work, AudioJudge~\citep{audiojudge}, evaluates the capabilities and limitations of using AudioLLMs for speech quality assessment and paralinguistic understanding via prompt engineering. Furthermore, many studies have focused on fine-tuning AudioLLMs to better expose their understanding capabilities for specific tasks, such as discriminating the human-likeness of audio~\citep{audio-turing-test}, modeling low-level acoustic qualities~\citep{chenchen-quality-audiollm,qualispeech}, unifying multiple speech quality evaluation tasks into a single AudioLLM~\citep{sq-llm}, and enhancing the assessment of instruction-following in spoken dialogue systems~\citep{wavreward,sagelm}. However, how to improve the ability of AudioLLMs to understand and judge speech {naturalness}, and how to use their quality-assessment capabilities as a reward to improve the post-training of speech generation models themselves, remain significantly underexplored.

\section{SpeechJudge-Data}


\begin{figure}[t]
\centering
\includegraphics[width=\textwidth]{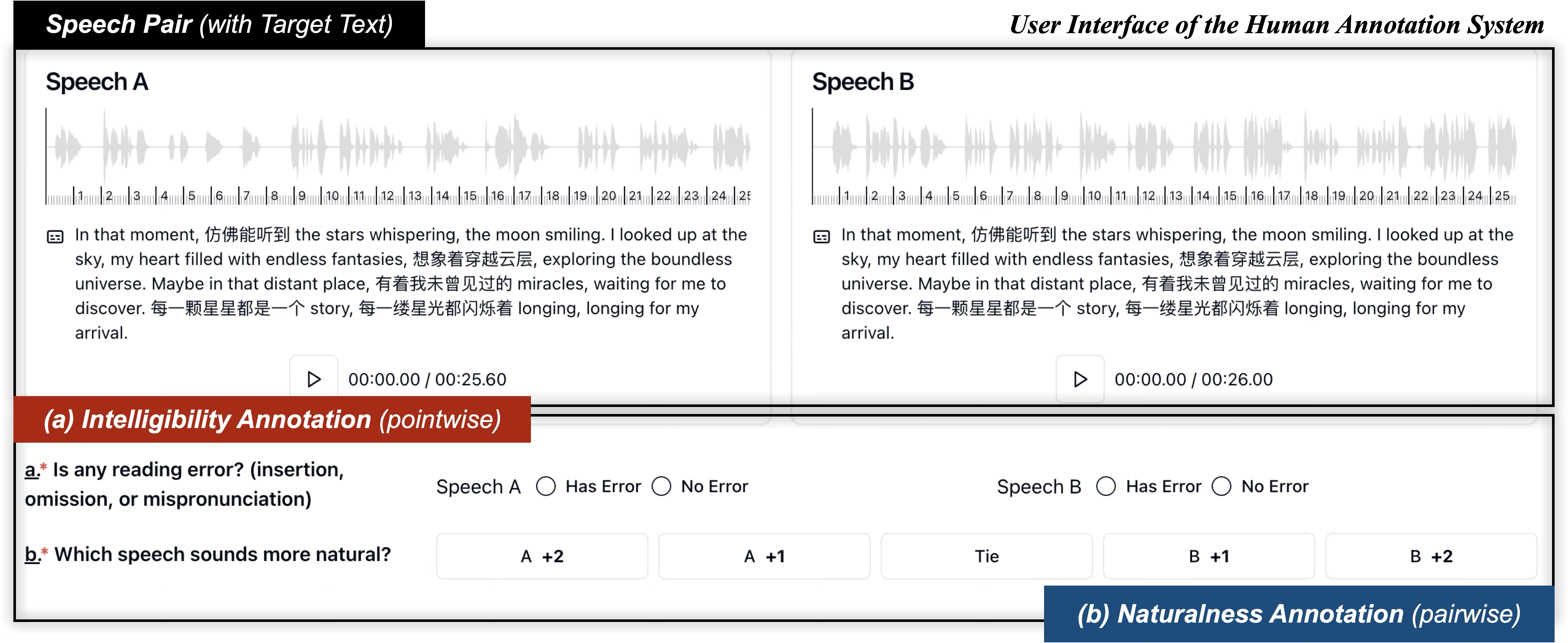}
\caption{SpeechJudge-Data consists of speech pairs (with corresponding text) synthesized by multiple zero-shot TTS models. For each pair, human annotators need to perform \textbf{\text{(a)}} a pointwise annotation of text accuracy to assess intelligibility, and \textbf{\text{(b)}} a pairwise preference annotation to judge the relative speech naturalness.}
\label{fig:system-ui}
\end{figure}

Our work is grounded in \textbf{SpeechJudge-Data}, a large-scale human feedback corpus for assessing the \text{\textit{intelligibility}} and \text{\textit{naturalness}} of synthesized speech. Formally, we aim to construct a dataset $\mathcal{D} = \{(t, a_1, a_2)\}$, where each triplet comprises a pair of synthesized speech samples $(a_1, a_2)$ and the corresponding target text $t$. We instruct annotators to provide pointwise intelligibility and pairwise naturalness preference annotations based on $\mathcal{D}$ (Figure~\ref{fig:system-ui}).

\subsection{Dataset Construction}
\label{sec:data-construction}

We employ a diverse set of recent advanced zero-shot TTS models to prepare the dataset $\mathcal{D}$. Formally, for each sample $(t, a_1, a_2)$, we denote the synthesized speech $a_i$ as being produced by the model $\mathcal{M}_{tts}$, i.e., $a_i \sim \mathcal{M}_{tts}(a_{ref}, t)$, where $a_{ref}$ is the reference speech. 


\begin{figure}[t]
    \centering 

    \begin{minipage}[b]{0.34\linewidth}
        \begin{subfigure}{\linewidth}
            \centering
            \includegraphics[width=\linewidth]{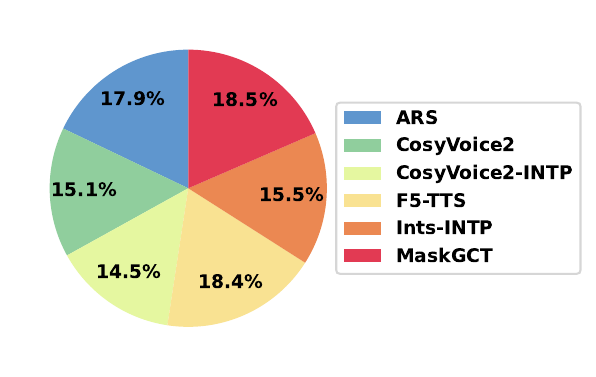}
            \caption{TTS Models.}
            \label{fig:distribution-model}
        \end{subfigure}
    \end{minipage}
    \hfill      
    %
    \begin{minipage}[b]{0.34\linewidth}
        \begin{subfigure}{\linewidth}
            \centering
            \includegraphics[width=\linewidth]{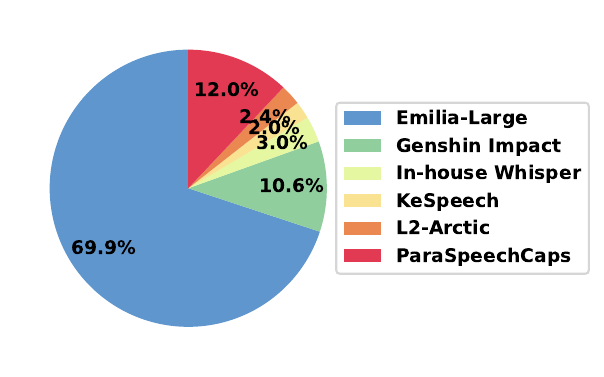}
            \caption{Speech References.}
            \label{fig:distribution-reference}
        \end{subfigure}
    \end{minipage}%
    \hfill
    %
    \begin{minipage}[b]{0.30\linewidth} 
        \begin{subfigure}{\linewidth}
            \centering
            \includegraphics[width=\linewidth]{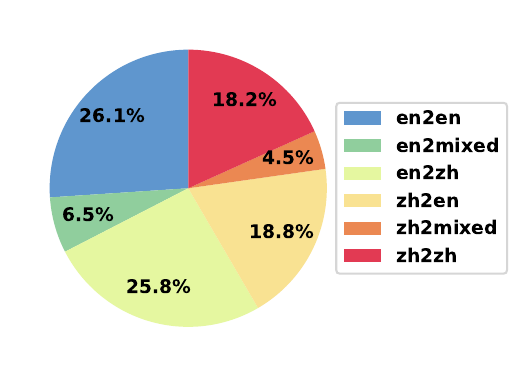}
            \caption{Language Settings.}
            \label{fig:distribution-language}
        \end{subfigure}
    \end{minipage}

    \caption{Distribution of SpeechJudge-Data.}
    \label{fig:distribution-speechjudge-data}
\end{figure}

\textbf{Model Selection}\quad
For $\mathcal{M}_{tts}$, we select the following six models of three architectures to enrich the distribution of the synthetic data (Figure~\ref{fig:distribution-model}): (1) \textbf{AR-based}: ARS~\citep{maskgct}, CosyVoice2~\citep{cosyvoice2}, CosyVoice2-INTP~\citep{intp}, and Ints-INTP~\citep{intp}. The latter two are released by~\citet{intp} as intelligibility-enhanced models. (2) \textbf{FM-based}: F5-TTS. (3) \textbf{MGM-based}: MaskGCT~\citep{maskgct}. 

\textbf{Prompt Construction}\quad
To build diverse prompts $(a_{ref}, t)$ for TTS, for $a_{ref}$, we adopt both \textbf{regular} and \textbf{expressive} speech samples. The regular samples are randomly selected from the Emilia-Large dataset \citep{emilia-large}. The expressive samples are sourced from corpora rich in paralinguistics, including the emotional corpora: ParaSpeechCaps~\citep{paraspeechcaps}, the accented corpora: L2-Arctic~\citep{l2arctic} and KeSpeech~\citep{kespeech}, the whisper samples from an in-house corpus, and the character voices from video games Genshin Impact~\citep{genshin}. We display the detailed distribution of speech references in Figure~\ref{fig:distribution-reference}.

The target text $t$ paired with each $a_{ref}$ is constructed as follows: For regular $a_{ref}$ samples, we randomly sample transcriptions from the Emilia-Large dataset~\citep{emilia-large}. These are then refined using DeepSeek-V3~\citep{deepseek-v3} to correct typos and normalize punctuations. For expressive $a_{ref}$ samples, we instruct DeepSeek-V3 to generate several scripts in different writing styles, tailored to the topic of $a_{ref}$ (see Appendix~\ref{app:deepseek-prompt} for more details). The languages of the target texts included Chinese (\textit{zh}), English (\textit{en}), and Chinese-English code-switching (\textit{mixed}). For the combinations $(a_{ref}, t)$, we include both monolingual settings (\textit{en2en} and \textit{zh2zh}) and cross-lingual settings (\textit{zh2en}, \textit{en2zh}, \textit{zh2mixed}, and \textit{en2mixed}), where \textit{zh2en} denotes Chinese $a_{ref}$ with English $t$, and similarly for others. The distribution of the language settings of $(a_{ref}, t)$ is shown in Figure~\ref{fig:distribution-language}.

\begin{figure}[ht]
    \centering
    \includegraphics[width=0.6\linewidth]{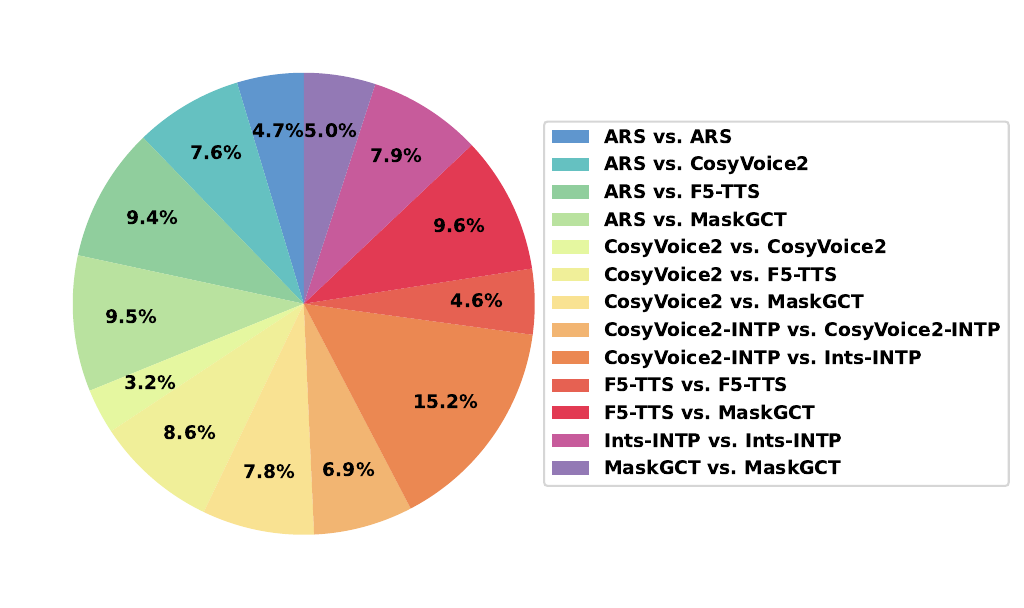}
    \caption{Distribution of the speech pairs of SpeechJudge-Data. The chart illustrates the percentage of both intra-model (e.g., ARS vs. ARS) and inter-model (e.g., ARS vs. CosyVoice2) pairs.}
    \label{fig:model_pairs_distribution}
\end{figure}

\textbf{Speech Pair Construction}\quad
To ensure the diversity of the $(a_1, a_2)$ pairs being compared, we follow~\citet{intp} and adopt both intra-model (i.e.,$a_1$ and $a_2$ being generated by the same model) and inter-model pairs (i.e., $a_1$ and $a_2$ being generated by the different models). The distribution of the speech pair is shown in Figure~\ref{fig:model_pairs_distribution}.

\subsection{Human Annotation}
Given a sample $(t, a_1, a_2)$, human annotators are instructed to perform both pointwise intelligibility and pairwise naturalness annotations (Figure \ref{fig:system-ui}). For intelligibility, annotators perform a binary classification to determine whether the speech ($a_1$ and $a_2$) accurately reads the text $t$ without any content insertion, omission, or mispronunciation. For naturalness, they perform a five-scale Comparative Mean Opinion Score (CMOS) annotation to determine which of the two audio clips ($a_1$ or $a_2$) sounds more natural and human-like. 

We recruited professional annotators from a specialized data annotation firm in China and provided them with training for speech naturalness judgement.  
All annotators assigned to Chinese data were native speakers. 
For the English and code-switching datasets, annotators were required to have a proficiency level equivalent to at least CET-6. All personnel underwent standardized training based on a detailed annotation manual. Initially, we conducted a pilot study among researchers to refine the guidelines for clarity and unambiguity. To ensure annotation quality, each sample $(t, a_1, a_2)$  was independently annotated by two individuals. A third annotator was introduced if any disagreements. The detailed annotation guidelines are provided in Appendix \ref{app:annotation}.

\textbf{Statistics}\quad
We recruit 69 annotators and conduct annotations over two months. The resulting constructed dataset $D$, which we denote as SpeechJudge-Data (raw), contains 99K $(t, a_1, a_2)$ samples, with each sample receiving an average of 2.49 annotations from different labelers. The market value of this annotation scale is estimated at over 500K RMB (about 70K USD). Based on the raw dataset, we also construct several subsets for analysis and reward model training. We provide detailed descriptions of each subset and its applications in the following sections and in Appendix~\ref{app:subsets-of-speechjudge-data}.


\textbf{Human Agreement Analysis}\quad
We analyze the human annotations for naturalness in this section; discussions regarding intelligibility are provided in Appendix~\ref{app:analysis-intelligibility}. For naturalness annotations, we evaluate the inter-annotator agreement across our constructed dataset. To simplify the analysis, given the sample $(t, a_1, a_2)$, we transform the five-scale naturalness scale (CMOS) into a ternary classification system: either $a_1$ is better, $a_2$ is better, or their quality is a Tie. Based on this simplified classification, we categorize the annotation results into four distinct levels of agreement\footnote{\textbf{Note}: Each sample of SpeechJudge-Data is independently annotated by a minimum of two and a maximum of three annotators (Appendix~\ref{app:annotation}).}: (1) \textbf{Full Agreement (FA)}: A consensus is reached among all annotators, with all ratings pointing to the same outcome (e.g., ``2A'', ``3A'', ``2B'', ``3B''). We use ``2A'' to indicate that two annotators both rated $a_1$ as better, while ``3B'' denotes three annotators all rating $a_2$ as better. (2) \textbf{Weak Agreement (WA)}: This level captures cases where two annotators agree on a specific polarity, while the third annotator marks a Tie (e.g., ``2A+1T'', ``2B+1T''). We also include the ``2T+1A'' and ``2T+1B'' cases in this level. (3) \textbf{Weak Disagreement (WD)}: This occurs when two annotators' ratings share the same polarity, but the third's rating is the opposite (e.g., ``2A+B'', ``2B+A''). (4) \textbf{Full Disagreement (FD)}: This represents a complete lack of consensus, where all three annotators provide different classifications, denoted as ``1A+1B+1T''.

\begin{figure}[ht]
    \centering
    \includegraphics[width=0.6\textwidth]{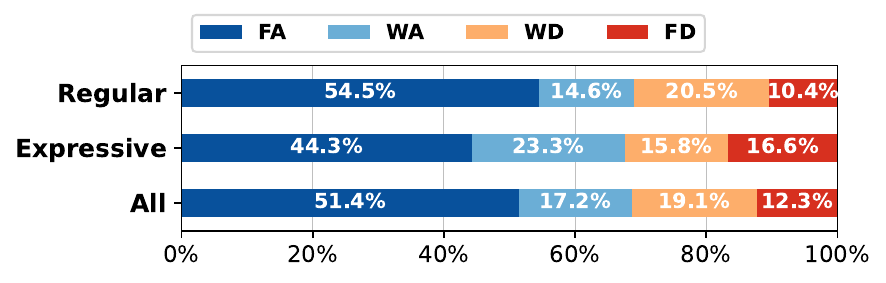}
    \vspace{-2mm}
    \caption{
    Distribution of SpeechJudge-Data on different levels of human agreement.
    }
    \label{fig:naturalness_agreement}
\end{figure}

In Figure~\ref{fig:naturalness_agreement}, we demonstrate the distribution of these human agreement levels for the SpeechJudge-Data and its two subsets, \textit{regular} and \textit{expressive} (which are defined by their speech references). The figure shows that about 70\% of the entire dataset falls into the Full Agreement (51.5\%) or Weak Agreement (17.2\%) levels. Furthermore, we observe that the \textit{expressive} subset has a lower agreement level than the \textit{regular} subset, which suggests that human evaluation of expressive speech generation is inherently a more challenging problem. Besides this sample-level agreement analysis, we also analyze the reliability of individual annotators, and we will discuss this in the Appendix~\ref{app:individual-annotator-analysis}.

\section{SpeechJudge-Eval}\label{sec:speechjudge-eval}
To evaluate speech naturalness, existing studies typically organize their own listening tests, which often have inconsistent settings across different papers~\citep{naturalspeech3,seedtts,cosyvoice,qwen2.5-omni,kimi-audio}. Alternatively, previous researchers use proxy MOS predictors, such as UTMOS~\citep{utmos}, as an objective metric. However, it remains an underexplored problem whether these metrics can accurately judge the naturalness of more advanced speech generation models~\citep{maskgct,cosyvoice2,f5tts,intp} and align with human preferences. Motivated by this, we construct a benchmark, \textbf{SpeechJudge-Eval}, specifically for the speech naturalness judgment task. 


\begin{figure}[t]
  \centering
  \begin{minipage}[t]{0.36\textwidth}
    \begin{table}[H]
    \caption{Protocols of different models for naturalness judgment.}
    \vspace{-3mm}
    \label{tab:eval-protocol}
    \centering
    \resizebox{\textwidth}{!}{%
    \begin{threeparttable}
        \begin{tabular}{cc}
            \toprule
             \multicolumn{1}{c}{\multirow{1}{*}{\textbf{Type}}} & \multicolumn{1}{c}{\textbf{Protocol}}  \\
            \midrule
            \multirow{3}{*}{\makecell[c]{\textbf{Objective}\\ \textbf{Metrics}}} & WER $\downarrow$, Naturalness $\uparrow$  \\
            &  SIM $\uparrow$, Naturalness $\uparrow$  \\
            &  FAD $\downarrow$, Naturalness $\uparrow$ \\
            \midrule
            \makecell[c]{\textbf{MOS}\\ \textbf{Predictors}} & MOS $\uparrow$, Naturalness $\uparrow$ \\
            \midrule
            \makecell[c]{\textbf{Deepfake}\\ \textbf{Detectors}} & \makecell[l]{Fake $\downarrow$, Naturalness $\uparrow$}  \\
            \midrule
            \textbf{AudioLLMs} & Score $\uparrow$, Naturalness $\uparrow$ 
        \end{tabular}
        \vspace{2mm}
        \begin{tabularx}{\linewidth}{X}
            \multicolumn{1}{c}{\textbf{\textit{Prompts of AudioLLMs}}} \\
            \vspace{-2mm}
            \setlength{\FrameSep}{3pt} 
            \begin{framed}
                \footnotesize
                \begin{itemize}[itemsep=0ex,leftmargin=2ex]
                \item We are comparing the naturalness of two  models' outputs. The models need to speak the target text accurately and naturally.
                \item Target text: \{$t$\}, Output A: \{$a_1$\}, Output B: \{$a_2$\}. Analyze the two outputs above, and score them with number from 1 to 10. Note:
                \vspace{-1mm}
                \begin{itemize}[label=$\circ$,itemsep=0ex,leftmargin=4ex]
                    \item  \textcolor{yushunblue}{Please evaluate the naturalness of both audio outputs based on the following criteria: Prosody and Intonation, Pacing and Rhythm, Articulation and Clarity, and Overall Naturalness.}
                    \item \textcolor{yushunblue}{After conducting a detailed analysis of each criterion,} using the following output template to highlight your conclusion: Output A: X, Output B: X.
                \end{itemize}
            \end{itemize}
            \end{framed}
            \\ \bottomrule
          \end{tabularx}
        \begin{tablenotes}
            \footnotesize{
                \item[*] We instruct AudioLLMs using two modes of prompt: \textit{plain} and \textit{CoT}. The text in \textcolor{yushunblue}{\text{blue}} is only employed during the \textit{CoT} mode.
            }
        \end{tablenotes}
    \end{threeparttable}    
    }   
    \end{table}
  \end{minipage}
  \hfill
  \begin{minipage}[t]{0.60\textwidth}
    \begin{table}[H]
    \caption{Accuracy of speech naturalness judgment across different models on SpeechJudge-Eval.}
    \vspace{-3mm}
    \label{tab:eval-benchmark}
    \centering
    \resizebox{\textwidth}{!}{%
    \begin{threeparttable}
        \begin{tabular}{lccc}
            \toprule
             \multicolumn{1}{c}{\multirow{1}{*}{\textbf{Model}}}  & \multicolumn{1}{c}{\textbf{Regular}} & \multicolumn{1}{c}{\textbf{Expressive}} & \multirow{1}{*}{\textbf{Total}} \\
            \midrule
            \rowcolor{gray!20}
            \multicolumn{4}{c}{\textit{\textbf{Objective Metrics}}} \\
            WER & 59.3 & 57.0 & 57.9 \\
            SIM & 47.5 & 42.5 & 44.5 \\
            FAD & 50.3 & 47.5 & 48.6 \\
            \midrule
            \rowcolor{gray!20} 
            \multicolumn{4}{c}{\textit{\textbf{MOS Predictor}}} \\
            {DNSMOS} &  61.0 & 55.8 & 57.9 \\
            {UTMOS} &  54.0 & 53.5 & 53.7 \\
            {Content Enjoyment (CE)} &  69.3 & 55.2 & 60.8 \\
            {Content Usefulness (CU)} &  61.3 & 54.7 & 57.3 \\
            {Production Complexity (PC)}  & 39.3 & 48.7 & 44.9 \\
            {Production Quality (PQ)} &  61.3 & 54.3 & 57.1 \\
            \midrule
            \rowcolor{gray!20} 
            \multicolumn{4}{c}{\textit{\textbf{Deepfake Detectors}}} \\
            AASIST & 40.5 & 50.8 & 46.7 \\
            ADV & 35.3 & 40.3 & 38.3 \\
            \midrule
            \rowcolor{gray!20} 
            \multicolumn{4}{c}{\textit{\textbf{AudioLLMs (Open-source)}}} \\
            {Phi-4-Multimodal} & 54.8 & 58.5 & 57.0 \\
            {Qwen2.5-Omni-7B} & 62.0 & 59.7 & 60.6 \\
            {Kimi-Audio-7B-Instruct} & 65.5 & 68.0 & 67.0 \\
            {Gemma-3n-E4B-it} & 49.0 & 47.7 & 48.2 \\
            {Voxtral-Mini-3B-2507} & 60.0 & 53.3 & 56.0 \\
            {MiDashengLM} & 58.8 & 63.5 & 61.6 \\
            MiMo-Audio-7B-Instruct & 61.3 & 49.3 & 54.1 \\
            \midrule
            \rowcolor{gray!20} 
            \multicolumn{4}{c}{\textit{\textbf{AudioLLMs (Closed-source)}}} \\
            {Gemini-2.5-Flash} & 73.5 & 66.2 & 69.1 \\
            {Gemini-2.5-Pro} & 73.0 & 62.2 & 66.5 \\
            {GPT-4o mini Audio} & 56.3 & 46.7 & 50.5 \\
            {GPT-4o Audio} & 71.5 & 64.7 & 67.4 \\
            \bottomrule
        \end{tabular}
        \begin{tablenotes}
            \footnotesize{
                \item[*] We use the protocols of Table~\ref{tab:eval-protocol} to establish judgment rules for different models. The results of AudioLLMs here are obtained using the \textit{plain} prompt of Table~\ref{tab:eval-protocol}. 
            }
        \end{tablenotes}
    \end{threeparttable}    
    }   
    \end{table}
  \end{minipage}
\end{figure}

\subsection{Task Description} 
\textbf{Task Formulation}\quad We formulate the naturalness judgment task as a pairwise comparison, specifically a \textit{win-or-lose} binary classification task: Given a target text $t$ and a corresponding audio pair $(a_1, a_2)$, a model needs to determine which audio has better naturalness. This results in a binary choice: either $a_1$ is better or $a_2$ is better. We use the human answer as the ground truth, and use \textbf{Accuracy} to measure the judgment performance of a model $\mathcal{M}$ on the evaluation set $\mathcal{D}$:
\begin{equation}
    \textbf{Accuracy} = \frac{1}{|\mathcal{D}|} \sum_{d=0}^{|\mathcal{D}|} \mathbb{I}(y_{\mathcal{M}}=y_{\mathcal{H}}),
\end{equation}
\label{eq:accuracy}
where $|\mathcal{D}|$ is the total number of samples in the evaluation set, $y_{\mathcal{M}}$ and $y_{\mathcal{H}}$ represent the answers of the model $\mathcal{M}$ and human for the sample $d$, respectively. $\mathbb{I}$ is the indicator function.

\textbf{Evaluation Data}\quad
We sample a subset of the SpeechJudge-Data to create the evaluation set for SpeechJudge-Eval. Specifically, we first select a subset that contains only \textit{preference data} (i.e., we filter out samples with the ``Tie'' annotation), and then choose only those with full-agreement-level (FA) samples to ensure a high-quality ground truth. We perform sampling from both the \textit{regular} and \textit{expressive} subsets of SpeechJudge-Data and proportionally cover the three target text languages (\textit{zh}, \textit{en}, and \textit{mixed}) within each subset. The final SpeechJudge-Eval dataset consists of 1,000 samples. The construction details of SpeechJudge-Eval and its distribution can be found in Appendix~\ref{app:subsets-of-speechjudge-data}.

\subsection{Benchmark for Different Models}

We test the naturalness judgment capability of various models based on SpeechJudge-Eval. We consider four different categories of models, whose evaluation protocols are shown in Table~\ref{tab:eval-protocol}:
\begin{enumerate}[itemsep=0ex,leftmargin=3ex]
    \item \textbf{Objective metrics}, such as WER~\citep{whisper,funasr}, SIM~\citep{wavlm}, and FAD~\citep{fad} in audio generation tasks. We assume that a better value of these metrics (e.g., lower for WER and FAD; higher for SIM) indicates better naturalness.
    \item \textbf{MOS Predictors}, including DNSMOS~\citep{dnsmos}, UTMOS~\citep{utmos}, and predictors from audiobox-aesthetics (CE, CU, PC, and PQ)~\citep{meta-audiobox-aesthetics}. We assume that a higher MOS score corresponds to better naturalness.
    \item \textbf{Deepfake detectors}, which are typically pre-trained on a binary classification task to predict whether an audio is fake or not~\citep{aasist,audio-deepfake-verification}. We assume that an audio with a lower fake probability should have better naturalness.
    \item \textbf{AudioLLMs}, which are employed to test their speech naturalness understanding capabilities in a \textit{zero-shot} manner\footnote{We assume that the adopted AudioLLMs have not been directly trained on the speech naturalness judgment task. Their performance on this benchmark is therefore considered a \textit{zero-shot} capability.}. We include the open-source Phi-4-Multimodal~\citep{phi4-mm}, Qwen2.5-Omni~\citep{qwen2.5-omni}, Kimi-Audio~\citep{kimi-audio}, Gemma-3n~\citep{gemma3}, Voxtral~\citep{voxtral}, MiDashengLM~\citep{midashenglm}, Mimo-Audio~\citep{mimoaudio}, and the closed-source Gemini-2.5~\citep{gemini2.5} and GPT-4o~\citep{gpt4o}. We use the \textit{plain} prompt of Table~\ref{tab:eval-protocol} to instruct the model to pairwise score the naturalness of two audios. We use their grading to determine the naturalness preference.
\end{enumerate}

The performance of different models on SpeechJudge-Eval is presented in Table~\ref{tab:eval-benchmark}. A key observation is that speech naturalness judgment is a highly challenging task. The leading model, Gemini-2.5-Flash, still only achieves less than 70\% agreement with human preferences. When comparing different models, we find that: (1) common objective metrics and MOS predictors show only a weak correlation with human preferences, often achieving less than 60\% accuracy and sometimes performing at the level of a random guess (around 50\%). (2) While deepfake detectors are highly effective at distinguishing between machine-generated and human-recorded speech~\citep{audio-deepfake-verification,aasist}, their ability to do so is not well-aligned with the naturalness objective when comparing two generated samples. (3) AudioLLMs demonstrate significant potential for this task. While some models, such as Gemma-3n and GPT-4o mini Audio, perform at a chance level, a number of others achieve an accuracy exceeding 60\%. This promising performance motivates us to further leverage these AudioLLMs for the design of a reward model for speech naturalness.


\section{SpeechJudge-GRM}\label{sec:speechjudge-grm}

\begin{figure}[t]
    \centering
    \includegraphics[width=\textwidth]{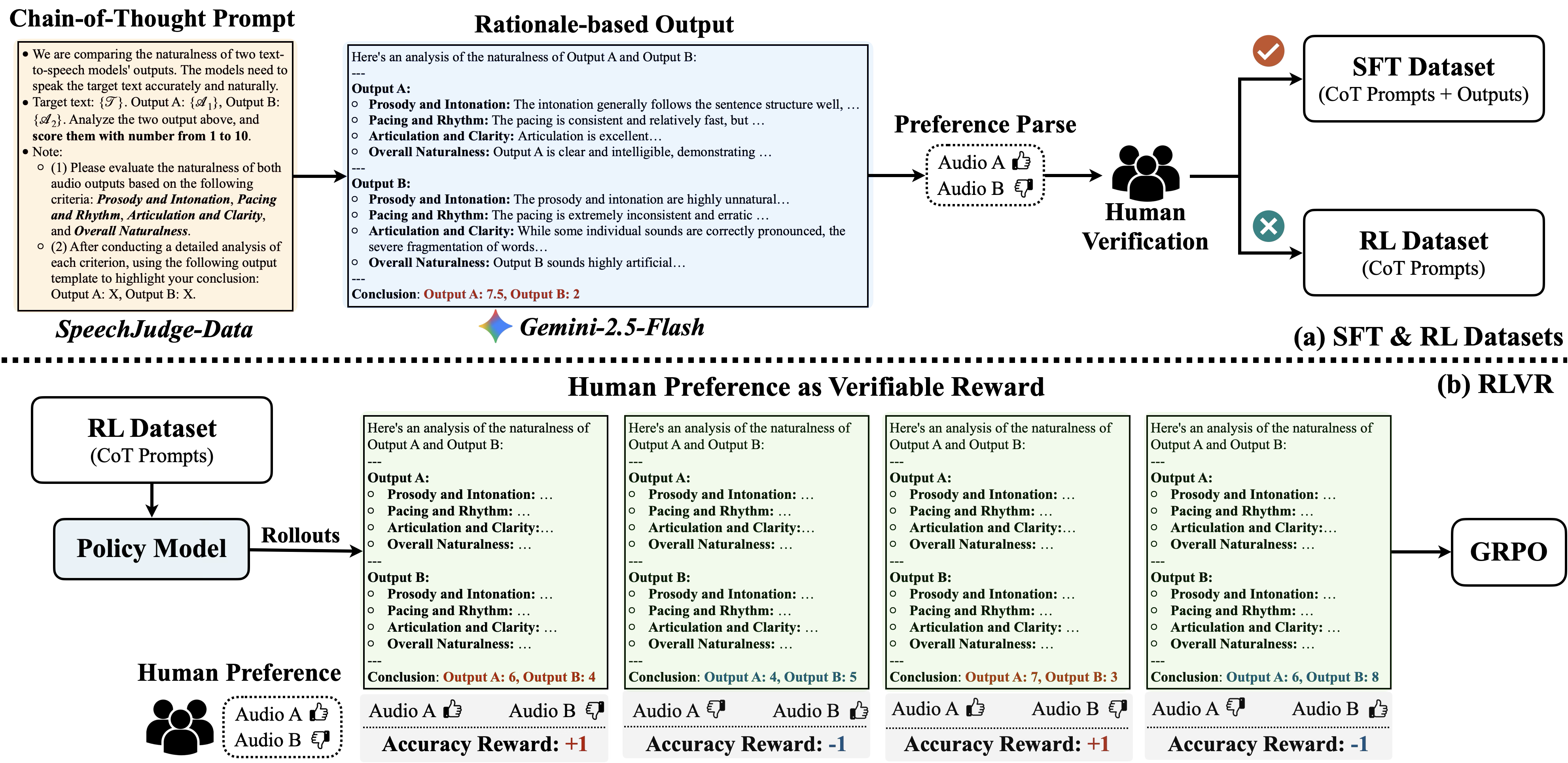}
    \caption{\textbf{SpeechJudge-GRM}: \text{(a)} We employ Gemini-2.5-Flash as a teacher model to generate CoT rationales for SpeechJudge-Data. We use the samples where Gemini-2.5-Flash's preference aligns with human as the SFT dataset, while the remaining samples are reserved for the RL stage. \text{(b)} We treat the human preference as a verifiable reward to train the GRM with GRPO.}
    \label{fig:speechjudge-grm}
\end{figure}

Based on the proposed SpeechJudge-Data, we further explore how to train a reward model capable of accurately capturing human preferences. Specifically, we propose \textbf{SpeechJudge-GRM}, where we leverage the inherent audio understanding capabilities of AudioLLMs (specifically, Qwen2.5-Omni-7B~\citep{qwen2.5-omni}) to elicit their speech naturalness judgment capability. Compared to the classic BTRM~\citep{bt-rm}, the key strengths of GRM are its ability to enable Chain-of-Thought (CoT) reasoning and its support for test-time computation via majority voting, which ultimately leads to improved preference judgment performance~\citep{google-grm}.



\subsection{Methodology}
We develop SpeechJudge-GRM based on Qwen2.5-Omni-7B (Thinker)~\citep{qwen2.5-omni}. Inspired by the powerful capabilities of RL with the verifiable reward (RLVR)~\citep{grpo,deepseek-r1}, our natural initial approach is to treat the human preference $y_{\mathcal{H}}$ for the pair $(a_1, a_2)$ as a verifiable reward, and launch a RLVR training based on Qwen2.5-Omni.
However, in practice, we find that the instruction-following reasoning capabilities of Qwen2.5-Omni are very weak (more detailed discussions can be found in Appendix~\ref{app:more-results-of-audiollm-benchmark}). Therefore, we adopt a two-stage post-training process (``SFT + RL'') to develop SpeechJudge-GRM (Figure~\ref{fig:speechjudge-grm}). We describe the details as follows.

\textbf{SFT Stage}\quad
We consider SFT as a ``cold start'' stage to improve the Qwen2.5-Omni's instruction-following, reasoning, and speech naturalness understanding capabilities. We select Gemini-2.5-Flash~\citep{gemini2.5}—one of the leading closed-source models on SpeechJudge-Eval (Table~\ref{tab:eval-benchmark})—to serve as a teacher model, and instruct it to generate the CoT data. Specifically, for each sample $d = (t, a_1,  a_2, y_{\mathcal{H}})$ from SpeechJudge-Data, we use the CoT prompt from Table~\ref{tab:eval-protocol} (denoted as $\mathbf{I}_{CoT}$) to instruct Gemini-2.5-Flash to generate a rationale-based output (denoted as $\mathbf{O}_{teacher}$). We then extract the preference judgment ($y_{\mathcal{M}}$) from this output. For samples where Gemini-2.5-Flash's preference is consistent with the human (i.e., $y_{\mathcal{M}} = y_{\mathcal{H}}$), we concatenate the CoT prompt and the model's output, $[\mathbf{I}_{CoT},\mathbf{O}_{teacher}]$, to create a data point for our SFT dataset. Conversely, we consider the sample $d$ a challenging case and reserve the prompt $\mathbf{I}_{CoT}$ for the second-stage RL dataset. During the SFT stage, for each training sample $[\mathbf{I}_{CoT},\mathbf{O}_{teacher}]$, we perform the next token prediction only on the segment $\mathbf{O}_{teacher}$.

\textbf{RL Stage}\quad
We treat the annotated human preference as a verifiable reward, and, building on the SFT model, we further trained it using the GRPO algorithm~\citep{grpo}. Specifically, for each sample $d = (t, a_1, a_2, y_{\mathcal{H}})$ in the RL dataset, we adopt the CoT prompt to instruct the policy model to conduct multiple rollouts during each iteration. For the $i$-th rollout, we parse the model's preference for $(a_1,  a_2)$, denoted as $y_{\mathcal{M}}^{i}$. Following~\citep{deepseek-grm}, we use an accuracy-based rule to calculate the reward: the reward is 1 if $y_{\mathcal{M}}^{i} = y_{\mathcal{H}}$, and -1 otherwise. In other words, during the RL stage, we only constrain the model's final naturalness judgment to align with human preferences, allowing the model to autonomously optimize its reasoning and rationale generation capabilities.

We denote the training dataset of SpeechJudge-GRM as SpeechJudge-Data (train). Its construction process is as follows (see Appendix~\ref{app:subsets-of-speechjudge-data} for more details). Based on the raw SpeechJudge-Data, we first filter out all samples at the Full Disagreement (FD) level. For the other samples—at the FA, WA, and WD levels—we apply a majority voting principle among annotators to determine the final label for each. We then further exclude samples with a ``Tie'' label, using only the remaining preference data to form the SpeechJudge-Data (train).
We use LoRA~\citep{lora} to fine-tune the GRM during both the SFT and RL stages. Other experimental setup details are provided in Appendix~\ref{app:expt-setup}.

\begin{figure}
  \centering
  \begin{minipage}[t]{0.52\textwidth}
    \begin{table}[H]
    \caption{Accuracy of speech naturalness judgment of SpeechJudge-GRM.}
    \vspace{-3mm}
    \label{tab:grm}
    \centering
    \resizebox{\textwidth}{!}{%
    \begin{threeparttable}
        \begin{tabular}{lccc}
            \toprule
             \multicolumn{1}{c}{\multirow{1}{*}{\textbf{Model}}} & \multicolumn{1}{c}{\textbf{Regular}} & \multicolumn{1}{c}{\textbf{Expressive}} & \multirow{1}{*}{\textbf{Total}} \\
            \midrule
        {Qwen2.5-Omni-7B} & 62.0 & 59.7 & 60.6 \\
        {Gemini-2.5-Flash} & 73.5 & 66.2 & 69.1 \\
        \midrule
        SpeechJudge-BTRM & 77.5 & 69.5 & 72.7 \\
        SpeechJudge-GRM \text{(SFT)} & 77.8 & 73.7 & 75.3  \\
        \multicolumn{1}{r}{w/ Voting@10} & 77.4 & 77.6 & 77.6 \\
        SpeechJudge-GRM \text{(SFT+RL)} & 79.0 & 76.0 & 77.2 \\
        \multicolumn{1}{r}{w/ Voting@10} & 80.5 & 78.7 & 79.4  \\
        \bottomrule
        \end{tabular}
        \begin{tablenotes}
            \footnotesize{\item[*] Our evaluation is conducted on SpeechJudge-Eval (like Table~\ref{tab:eval-benchmark}). \textbf{w/ Voting@10}: For each prompt, the GRM generates 10 outputs, and we use the majority voting from these 10 outputs as the final result.}
        \end{tablenotes}
    \end{threeparttable}    
    }   
    \end{table}
  \end{minipage}
  \hfill
  \begin{minipage}[t]{0.45\textwidth}
    \centering
    \begin{figure}[H]
        \includegraphics[width=\linewidth]{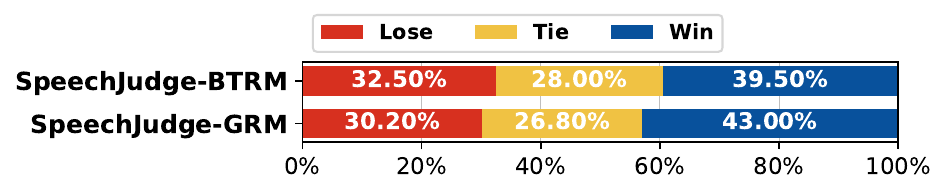}
        \caption{
        Subjective evaluation of using SpeechJudge-GRM for high-naturalness sample selection. Human subjects compare a best-of-100 output of Qwen2.5-Omni-7B (Talker), chosen by either SpeechJudge-BTRM or SpeechJudge-GRM, against a randomly output.
        }
        \label{fig:best-of-n-by-speechjudge-grm}
    \end{figure}
  \end{minipage}
\end{figure}

\subsection{Effectiveness of SpeechJudge-GRM on Naturalness Judgement}\label{sec:grm-results}

To verify the effectiveness of SpeechJudge-GRM for naturalness judgment, we evaluate it on the SpeechJudge-Eval benchmark. We develop SpeechJudge-BTRM as a baseline, which utilizes the BTRM paradigm~\citep{bt-rm,dpo} by adding a linear layer on Qwen2.5-Omni-7B (Thinker) to produce a single scalar reward prediction. SpeechJudge-BTRM also uses LoRA fine-tuning and uses the same training data as SpeechJudge-GRM.

From the results of Table~\ref{tab:grm}, we can observe that: (1) The SpeechJudge-BTRM achieves a 72.7\% agreement with human preferences on SpeechJudge-Eval, a level of performance comparable to the initial development of BTRMs in the textual LLM RLHF field~\citep{openai-rm,RLHF-anthoropic,instructgpt}. (2) After conducting SFT training with the CoT data, the accuracy of SpeechJudge-GRM (SFT) reaches 75.3\%. Besides, further RLVR training improves the final model SpeechJudge-GRM (SFT+RL) to an accuracy of 77.2\%. (3) Due to the generative nature of the GRM, we can further enhance the accuracy of SpeechJudge-GRM using inference-time scaling. For example, by using majority voting across 10 outputs instead of just one, the accuracy is improved by approximately 2 percentage points (75.3\% $\rightarrow$ 77.6\%; 77.2\% $\rightarrow$ 79.4\%). These results collectively verify the effectiveness of our proposed SpeechJudge-GRM for judging speech naturalness.

\subsection{High-Quality Sample Selection based on SpeechJudge-GRM}\label{sec:sample-selection}

We investigate the effect of SpeechJudge-based reward models for high-quality sample selection. We use the hard cases from SeedTTS-Eval~\citep{seedtts} and the code-switching cases from Amphion-TTS-Eval~\citep{amphion} as target texts. For each text, we instruct the Qwen2.5-Omni-7B (Talker)~\citep{qwen2.5} to generate 100 speeches. We then ask human subjects to compare the best-of-100 output—as selected by either SpeechJudge-BTRM or SpeechJudge-GRM—against a randomly sampled output. The evaluation measures the win/lose/tie ratios based on speech naturalness. From Figure~\ref{fig:best-of-n-by-speechjudge-grm}, we observe that the best-of-100 samples selected by both SpeechJudge-BTRM and SpeechJudge-GRM are more likely to outperform a randomly selected sample from the same set. This finding demonstrates the advantage of using the SpeechJudge-Data corpus for training human-aligned reward model. Furthermore, SpeechJudge-GRM exhibits better performance than SpeechJudge-BTRM, which highlights the superiority of the proposed GRM.

\subsection{Post-Training of Zero-Shot TTS based on SpeechJudge-GRM}\label{sec:post-training-for-tts}

We investigate the effect of using SpeechJudge-GRM as a reward function for post-training of TTS model. Specifically, we develop a new zero-shot TTS model, \textbf{Qwen2.5-0.5B-TTS}, to serve as the base model, which was not involved in the construction of the SpeechJudge-Data. This model is based on Qwen2.5-0.5B~\citep{qwen2.5}, adopts the classic two-stage ``AR+Diffusion'' architecture~\citep{seedtts,cosyvoice}, uses the speech tokenizer from DualCodec~\citep{dualcodec}, and is pre-trained on the Emilia dataset~\citep{emilia-large}.

\begin{figure}[t]
    \centering
    \begin{subfigure}[c]{0.44\textwidth}
    \centering
    \resizebox{\linewidth}{!}{%
        \begin{tabular}{lrcc}
            \toprule
            \multicolumn{1}{c}{\textbf{Model}} & \textbf{T-ACC} & \textbf{N-CMOS} \\
            \midrule
            \multicolumn{1}{l}{\text{Qwen2.5-0.5B-TTS}} & 84.0\% & 0.00 \textcolor{white}{$_{\scriptscriptstyle \pm \text{0.00}}$} \\
            \midrule
            \multicolumn{1}{r}{\text{w/ INTP}} & 87.0\% & 0.18 $_{\scriptscriptstyle \pm \text{0.07}}$ \\
            \multicolumn{1}{r}{\text{w/ SpeechJudge-Data}} &  91.0\% & 0.16 $_{\scriptscriptstyle \pm \text{0.08}}$ \\
            \multicolumn{1}{r}{\text{w/ SpeechJudge-GRM (offline)}} & 91.0\% & 0.21 $_{\scriptscriptstyle \pm \text{0.12}}$ \\
            \multicolumn{1}{r}{\text{w/ SpeechJudge-GRM (online)}} & 90.0\% & 0.25 $_{\scriptscriptstyle \pm \text{0.09}}$ \\
            \bottomrule
        \end{tabular}%
    }
    \caption{Text Accuracy (T-ACC) and Naturalness CMOS (N-CMOS).}
    \label{fig:post-train-tts-naturalness}
    \end{subfigure}
    \hfill
    \begin{subfigure}[c]{0.53\textwidth}
        \centering
        \vspace{-4mm}
        \includegraphics[width=\linewidth]{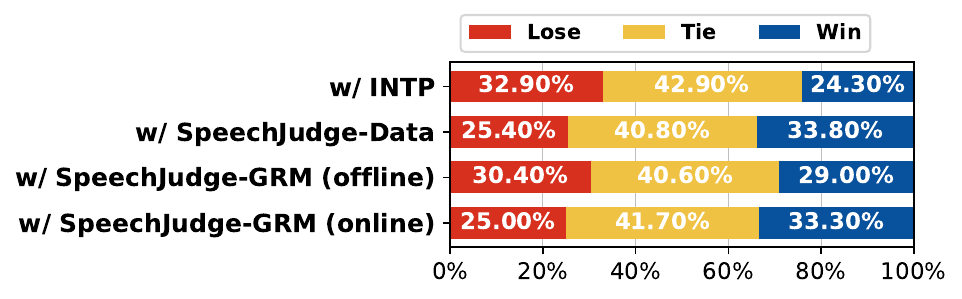}
        \vspace{-2mm}
        \caption{Win/Lose/Tie of speaker similarity after post-training.}
        \label{fig:post-train-tts-speaker-similarity}
    \end{subfigure}
    \caption{Post-training of Qwen2.5-0.5B-TTS based on SpeechJudge. We display the objective results (WER and SIM) in Table~\ref{tab:grm-post-trained-tts-details} of Appendix~\ref{app:post-training-tts}.}
    \label{fig:post-training-tts-results}
\end{figure}

Based on this pre-trained model, we design four comparative methods:
(1) \textbf{w/ INTP}: We use the intelligibility preference dataset, INTP~\citep{intp}, to perform offline DPO alignment~\citep{dpo}.
(2) \textbf{w/ SpeechJudge-Data}: We use the SpeechJudge-Data (train) to perform offline DPO alignment.
(3) \textbf{w/ SpeechJudge-GRM (offline)}: We use SpeechJudge-GRM as an offline preference data annotator. We take all speech pairs from the INTP dataset and re-annotate their preference labels using SpeechJudge-GRM, then perform offline DPO alignment on the resulting data.
(4) \textbf{w/ SpeechJudge-GRM (online)}: We use SpeechJudge-GRM as a reward function for the online DPO algorithm~\citep{online-dpo}. The training data consists of only the prompts from INTP (i.e., the target texts and speech references for zero-shot TTS).

We use SeedTTS-Eval~\citep{seedtts} and Amphion-TTS-Eval~\citep{intp,amphion,vevo2} as evaluation sets. We present the objective results (WER and SIM) in Table~\ref{tab:grm-post-trained-tts-details} and the subjective results in Figure~\ref{fig:post-training-tts-results}. 
We observe that both intelligibility and naturalness are enhanced for all the four methods after post-training. Additionally, the post-training method based on SpeechJudge-GRM achieves a greater improvement in naturalness (Figure~\ref{fig:post-train-tts-naturalness}). Besides, the SpeechJudge-based methods could match or lead to a slight improvement in speaker similarity (Figure~\ref{fig:post-train-tts-speaker-similarity}).

\section{Limitations and Future Work}
\label{sec:limitations}

While SpeechJudge-Data and SpeechJudge-GRM represent a step toward
human-aligned speech naturalness judges, several limitations remain and open
up directions for future work.

\textbf{Scope of data and annotators.}\quad
Our corpus is constructed entirely from synthetic TTS outputs in Chinese,
English, and Chinese--English code-switching, and our annotators are
professional raters in China (native Mandarin speakers with high but still L2
English proficiency).
As shown in Appendix~\ref{app:inter-annotator}, inter-annotator agreement is noticeably higher on
Chinese than on English and mixed subsets, indicating that the current
dataset primarily reflects the preferences of Chinese and Chinese--English
bilingual listeners, and is tailored to TTS-style read speech rather than
spontaneous conversation.
Extending SpeechJudge-Data to more languages, speaking styles, and listener
populations (including native speakers of other languages and more diverse
cultural backgrounds) is an important direction for building more universal
naturalness judges.

\textbf{Residual failure cases.}\quad
Our error analysis in Appendix~\ref{app:grm-error} shows that SpeechJudge-GRM’s remaining
mistakes on {SpeechJudge-Eval} concentrate on
some specific trade-offs, such as clean but robotic vs.\ slightly noisy but lively speech,
prosody vs.\ articulation, and extreme expressive styles like very
high-F0 emotional speech or whispers.
In these regimes the model can over-weight cleanness, under-weight
style-appropriate prosody, or become effectively indifferent when preference
gaps are extremely small.
Future work could incorporate explicit modeling of recording conditions
(e.g., background noise), style-aware priors, and targeted augmentation of
expressive and cross-lingual examples to better capture these nuanced
preferences.

\textbf{CoT quality and teacher bias.}\quad
The CoT capability of SpeechJudge-GRM is bootstrapped from a proprietary
teacher (Gemini-2.5-Flash) via an SFT stage.
Although our analyses in Appendix~\ref{app:grm-cot-quality} suggest that GRM’s CoT is largely
self-consistent and moderately faithful,
the reasoning style still
inherits biases from the teacher, and we do not perform large-scale human
verification of the intermediate explanations.
An interesting direction is to involve humans more directly in assessing and curating CoT rationales—similar in spirit to the concurrent SQ-LLM work~\citep{sq-llm}, which incorporates human involvement in CoT annotation—for example, by collecting human-written or human-edited analyses, or learning from explicit feedback on explanation quality, and to explore alternative, fully open-source teachers for bootstrapping.

\textbf{From coarse-grained to fine-grained naturalness.}\quad
In this work, naturalness is annotated and modeled at the \textit{utterance}
level: for each sentence, annotators choose which of two speeches is more
natural, and SpeechJudge-GRM outputs a single decision per pair.
However, in real-world speech, naturalness is often highly non-uniform within
an utterance—some segments sound very natural while others contain local
artifacts, disfluencies, or prosodic issues.
Our current formulation does not explicitly localize such fine-grained
phenomena.
A promising future direction is to collect segment-level or time-aligned
human feedback and to train reward models that can produce not only
utterance-level judgments but also fine-grained scores or rationales over
time. 

\section{Conclusion}

In this work, we tackle the challenge of aligning speech synthesis with human perception of naturalness by introducing SpeechJudge: a suite consisting of a large-scale human preference dataset (SpeechJudge-Data), a challenging benchmark (SpeechJudge-Eval), and a generative reward model (SpeechJudge-GRM). Our benchmark shows that even strong AudioLLMs struggle at naturalness judgment, reaching under 70\% agreement with human preferences. In contrast, the proposed SpeechJudge-GRM achieves 77.2\% accuracy on SpeechJudge-Eval (up to 79.4\% with inference-time scaling @10), outperforming a classic Bradley–Terry reward model (72.7\%). We further demonstrate that SpeechJudge-GRM serves as an effective reward function for post-training TTS models, leading to improved perceived naturalness in downstream evaluations. By releasing our data, benchmark, and models, we hope to enable further research on human-aligned speech generation and more reliable evaluation of speech naturalness.


\section*{Ethics Statement}
Our dataset was constructed with feedback from paid professional annotators under fair labor conditions, and the data itself consists of synthesized speech from properly licensed corpora, safeguarding the privacy of all individuals. {All human annotations were collected from professional annotators recruited by a third-party data annotation company under written informed consent, on synthetic TTS audio only, with no collection of personally identifying information.} We acknowledge that our models may reflect linguistic biases present in the English and Chinese source data and recognize that generative speech technology has dual-use potential. We do not condone any malicious use of our work, such as the creation of misleading deepfakes.

\bibliographystyle{plainnat}
\bibliography{main}


\beginappendix

\section{Details of SpeechJudge-Data}\label{app:speechjudge-data}

\subsection{Details of Prompt Construction}
\label{app:deepseek-prompt}

For the target texts paired with the regular speech references, we use DeepSeek-V3~\citep{deepseek-v3} to fix typos and normalize punctuations, the prompt used is listed below.

\begin{promptbox}
\textit{\textbf{System Prompt:} }\\
I obtained a text from an audio file based on some ASR models. Please help me clean it up (e.g., correct typos, add proper punctuation marks, and make the sentences semantically coherent). Note: (1) You can modify, add, or replace words that better fit the context to ensure semantic coherence. (2) Please only return the cleaned-up result without any explanation.

\medskip \textit{\textbf{User Prompt (Example):} }\\
a panda eats shoes and leaves

\medskip \textit{\textbf{System Output (Example):} }\\
A panda eats shoots and leaves.
\end{promptbox}

For the target texts paired with the expressive speech references, we use DeepSeek-V3 to generate several scripts in different writing styles based on the speech reference's text, the prompt used is listed below.

\begin{promptbox}
\textit{\textbf{System Prompt:} }\\
\begin{CJK*}{UTF8}{gbsn}
    请按句子原始语言进行续写（中文：续写中文，英文：续写英文）。将给出的句子以10种截然不同的文风续写（中文：如古典文学风格、科幻风格、武侠风格等；英文：Formal, Informal, Literary, Journalistic, Advertising, etc.），直接续写即可，续写时避免重复原句内容。每段续写文本长度应控制在正常语速朗读约10-20秒左右。

    请严格按照以下JSON格式返回结果，不添加任何其他解释或备注：\\
    \{\\
    "continue": [ \\
    \{"type": "文风类型", "content": "续写内容"\}, \\
    \{"type": "文风类型", "content": "续写内容"\}, \\
    ...（共10条） \\
    ], \\
    "random": "从以上续写中随机选出一条续写内容", \\
    "translation": "将random内容经过中英互译处理后的文本", \\
    "codeSwitching": "将random内容进行适度的中英code-switching，即在句中选取部分关键词汇转换为英文或中文后的文本"\\
    \}
\end{CJK*}

\medskip \textit{\textbf{User Prompt (Example):} }\\
A panda eats shoots and leaves.

\medskip \textit{\textbf{System Output (Example):} }\\
\begin{CJK*}{UTF8}{gbsn}
\{\\
"continue": [\\
\{"type": "Formal", "content": "This behavior is characteristic of the species Ailuropoda melanoleuca, which sustains itself primarily on a herbivorous diet consisting of various bamboo species. The consumption of both young shoots and mature leaves is essential for meeting its nutritional requirements."\},\\
\{"type": "Informal", "content": "And then, after chowing down on all that bamboo, it just kinda chills out, maybe takes a long nap or something. They're pretty laid-back animals, you know? Not much else going on when you're full of leaves."\},\\
...\\
\{"type": "Fantasy", "content": "Legend says that with every shoot consumed, a little more of the forest's ancient magic is absorbed into its being. The leaves it leaves behind are not mere waste, but enchanted tokens that can heal or curse, depending on the heart of the finder."\},\\
...\\
],\\
"random": "Legend says that with every shoot consumed, a little more of the forest's ancient magic is absorbed into its being. The leaves it leaves behind are not mere waste, but enchanted tokens that can heal or curse, depending on the heart of the finder.",\\
"translation": "传说中，它每吃下一根竹笋，就会吸收一点森林的古老魔力。它身后留下的叶子并非单纯的废弃物，而是附有魔法的信物，可治愈亦可诅咒，全凭发现者的心地。",\\
"codeSwitching": "Legend says that with every shoot consumed, a little more of the forest's ancient magic 被吸收 into its being. The leaves it leaves behind are not mere waste, but enchanted tokens that can heal 或 curse, depending on the finder 的 heart."\\
\}
\end{CJK*}

\end{promptbox}

\subsection{Subsets of SpeechJudge-Data}\label{app:subsets-of-speechjudge-data}


\begin{figure}
    \centering
    \includegraphics[width=0.9\linewidth]{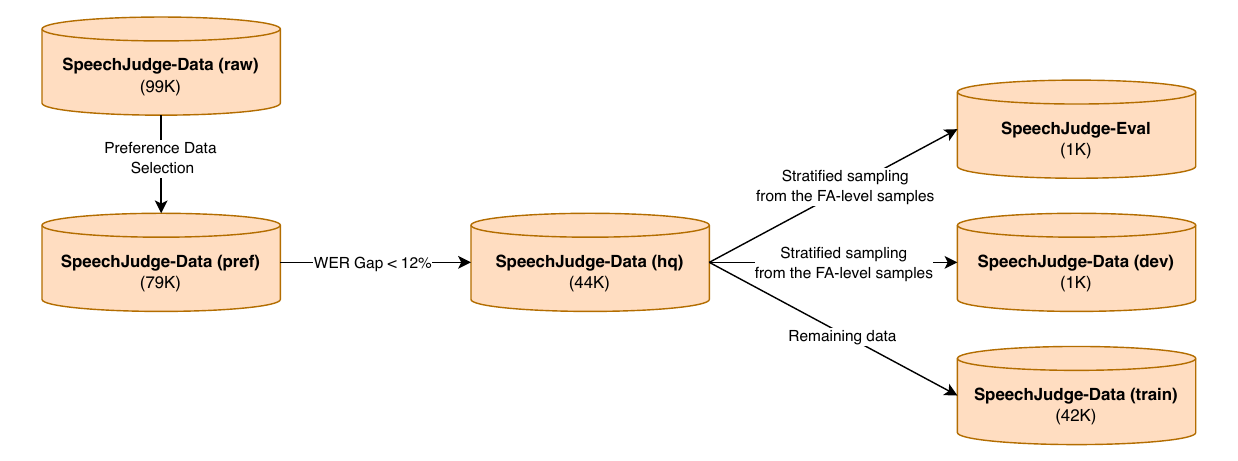}
    \caption{From the raw SpeechJudge-Data to its different subsets.}
    \label{fig:data-construction}
\end{figure}


We construct several subsets based on SpeechJudge-Data (Figure~\ref{fig:data-construction}). We begin with the \textbf{SpeechJudge-Data (raw)} corpus, containing 99K pairs, where each pair is annotated by multiple labelers as a five-scale naturalness CMOS. We aggregate these annotations via a majority vote for each pair, and subsequently discard all ``Tie'' pairs, yielding the 79K-pair human \textit{preference} data, denoted as \textbf{SpeechJudge-Data (pref)}.

During our preliminary analysis based on SpeechJudge-Data (pref), we observe that a significant disparity in intelligibility between two speech samples can overshadow the subtler quality of naturalness, biasing human preference toward the more comprehensible sample. To mitigate this confounding factor and create a more high-quality dataset focused specifically on naturalness, we further refine the data. Specifically, we retain only pairs where the absolute WER gap of those is below 12\%. This process results in the 44K-pair high-quality \textbf{SpeechJudge-Data (hq)} subset, ensuring that its preference labels are more reflective of genuine differences in naturalness.

From SpeechJudge-Data (hq), we construct our benchmark, \textbf{SpeechJudge-Eval}, by applying stratified sampling to FA-level pairs, resulting in 1,000 pairs; its composition is detailed in Table~\ref{tab:speechjudge-eval-data}. Similarly, we use the same strategy to construct a validation set of the same size, \textbf{SpeechJudge-Data (dev)}. The remaining 42K pairs, \textbf{SpeechJudge-Data (train)}, constitute the training set for our reward models.

\begin{table}[ht]
\centering
\caption{Distribution of the {{SpeechJudge-Eval}} benchmark.}
\label{tab:speechjudge-eval-data}
\resizebox{0.8\textwidth}{!}{
\begin{tabular}{lccc}
\toprule
\textbf{Subset} & \textbf{Source of Speech References} & \textbf{Languages of Target Texts} & \textbf{\# Pairs} \\ \midrule
\multirow{2}{*}{\textbf{Regular}} & \multirow{2}{*}{Emilia-Large} & \textit{en} & 200 \\
 & & \textit{zh} & 200 \\
 \midrule
\multirow{3}{*}{\textbf{Expressive}} & \multirow{3}{*}{\makecell[c]{ParaSpeechCaps, L2-Arctic,\\KeSpeech, Genshin, etc.}} & \textit{en} & 200 \\
 & & \textit{zh} & 200 \\
 & & \textit{mixed} & 200 \\
\bottomrule
\end{tabular}
}
\end{table}

\section{Human Annotation Details}
\label{app:annotation}

The complete annotation guidelines are attached below:

\begin{promptbox}
\subsection*{Task Introdcution}

In each task, you are required to complete two evaluations:
\begin{enumerate}
    \item \textbf{Pronunciation Error Detection} \\
    For each audio clip, we provide the \textbf{target text} that is intended to be read aloud. You need to determine if there are any pronunciation errors in the audio, such as omissions (missing words), insertions (extra words), or substitutions (wrong words).
    
    \item \textbf{Naturalness Comparison} \\
    Compare two audio clips and determine which one sounds more natural and more like a real human speaking.
\end{enumerate}

\textbf{Attention:}
\begin{itemize}
    \item Please use headphones for the evaluation to better capture audio details and improve judgment accuracy.
\end{itemize}

\subsection*{Annotation Criteria}
\subsubsection*{a. Pronunciation Error Detection}
Pronunciation errors include the following three categories:
\begin{itemize}
    \item \textbf{Omission:} Certain words from the target text are missed.
    \item \textbf{Insertion:} Extra words not in the target text are added, e.g., repeating words.
    \item \textbf{Substitution:} Certain words are misread, e.g., reading names, numbers, polyphonic characters, or other words incorrectly, or making word order errors.
\end{itemize}

\textbf{Attention:} These pronunciation errors can occur at any point in the audio. For example:
\begin{itemize}
    \item At the beginning of the audio, words are spoken that are not in the target text.
    \item At the end of the audio, some content from the target text is omitted.
    \item In the middle of the audio, omissions, insertions, or substitutions occur.
    \item \dots
\end{itemize}

\subsubsection*{b. Naturalness Comparison}
Natural speech should sound like a real person talking. Specific criteria include:
\begin{itemize}
    \item The audio is clear, free from robotic/electronic tones or obvious noise (e.g., unnatural laughter, shouting, or other irrelevant background voices).
    \item The intonation is natural and expressive, not flat or mechanical.
    \item The speaking rhythm is reasonable, with moderate speed and appropriate pauses. (Note: Inappropriate pauses affect the naturalness of the audio but are not classified as ``pronunciation errors'').
    \item Word stress is placed correctly, conforms to normal linguistic sense, and does not sound abrupt.
\end{itemize}

The rating scale is as follows:
\begin{itemize}
    \item \textbf{A +2:} Audio A is significantly more natural than B (large difference).
    \item \textbf{A +1:} Audio A is slightly more natural than B (slight difference).
    \item \textbf{Tie:} The naturalness of the two audio clips is similar and difficult to judge.
    \item \textbf{B +1:} Audio B is slightly more natural than A (slight difference).
    \item \textbf{B +2:} Audio B is significantly more natural than A (large difference).
\end{itemize}

\textbf{Attention:}
\begin{itemize}
    \item If there are minor errors in individual words, please do not let them affect your overall judgment of naturalness.
    \item However, if a large number of content errors are found that severely interfere with your listening experience, this will affect the audio's naturalness.
\end{itemize}

\end{promptbox}

\subsection{Individual Annotator Reliability}\label{app:individual-annotator-analysis}
To assess the reliability of individual annotators, we computed agreement rate for each participant. This rate measures the extent to which an annotator's judgments align with those of their peers on the same sample $(t, a_1, a_2)$.

For a given sample annotated by a group of $M$ annotators, the agreement score for annotator $i$ is calculated as the fraction of the other $M-1$ annotators who assigned the exact same label. An annotator's final reliability score is the average of these scores across all samples they evaluated. We excluded participants who annotated fewer than 10 samples from this analysis.

Formally, for an annotator $i$ who labeled $N_i$ samples, the agreement rate $r_{ij}$ for sample $j$ is defined as:
\[
r_{ij} \;=\; \frac{1}{M-1} \sum_{x \neq i} \mathbb{I}\!\left[y_{xj} = y_{ij}\right]
\]
The overall agreement rate for annotator $i$, denoted as $R_i$, is then:
\[
R_i \;=\; \frac{1}{N_i} \sum_{j=1}^{N_i} r_{ij}
\]
where $y_{ij} \in \{A, B, T\}$ is the label assigned by annotator $i$ to sample $j$. The label $T$ (i.e., Tie) is treated as a distinct category, and it's agreement is counted only on exact matches.

\begin{figure}[ht]
    \centering
    \includegraphics[width=\textwidth]{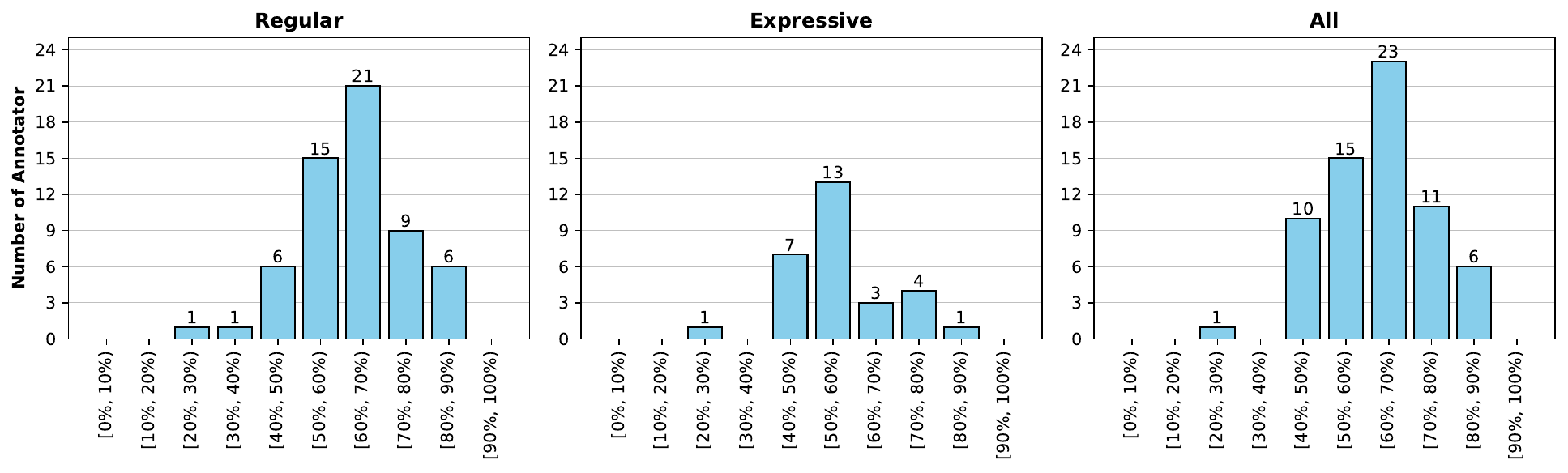}
    \caption{Distribution of individual annotator agreement rates. Most annotators fall between 50\% and 80\%, typically peaking in the 60--70\% range.}
    \label{fig:rater_agreement_dist}
\end{figure}

Figure~\ref{fig:rater_agreement_dist} illustrates the distribution of these agreement rates for our 69 annotators for SpeechJudge-Data (raw). The distribution is generally unimodal with a peak in the 60--70\% range\footnote{We have noted that one annotator's agreement with the others is less than 30\%, so we ultimately removed his data from SpeechJudge-Data.}, which indicates a consistent and reliable level of performance across the annotation pool.

\subsection{Inter-Annotator Agreement}
\label{app:inter-annotator}

Complementary to the per-annotator reliability analysis in
Appendix~\ref{app:individual-annotator-analysis}, we now quantify the overall level of
agreement among annotators at the dataset level.
Following common practice in RLHF-style preference datasets, we compute inter-annotator agreement as the probability that two annotators chosen at random assign the same preference label to the same pair~\citep{openai-rm,instructgpt,imagereward}.
Table~\ref{tab:inter-agreement} summarizes the results.

\textbf{Ternary vs.\ binary preferences.}\quad
\text{SpeechJudge-Data (raw)} contains ternary labels
(i.e., {A better} / {B better} / {Tie}), and the corresponding
inter-annotator agreement is $50.7\%\pm0.1\%$ over the whole corpus.
After removing ``Tie'' cases and restricting to clear binary preferences
(\text{SpeechJudge-Data (pref)}, labels
{A better} / {B better} only), the agreement rises to
$69.0\%\pm0.2\%$.
This level is comparable to well-established RLHF datasets in text and vision:
ImageReward reports an agreement of $65.3\%\pm1.6\%$~\citep{imagereward},
and InstructGPT reports $72.6\%\pm1.5\%$ on binary A/B comparisons~\citep{instructgpt}.
These numbers indicate that, despite the inherent subjectivity of the task,
our binary preference data are on par with existing human-feedback corpora.

\textbf{Effect of style.}\quad
Within each language, expressive prompts are slightly harder than regular ones:
for example, on \text{SpeechJudge-Data (pref)}, regular \textit{zh} reaches
$74.4\%\pm0.2\%$ agreement while expressive \textit{zh} reaches
$73.0\%\pm0.6\%$.
A similar, mildly lower agreement is observed for expressive \textit{en} and
\textit{mixed} subsets.
Overall, however, regular vs.\ expressive speech remains in a similar agreement
range, suggesting that our guidelines allow annotators to make consistent naturalness judgments.

\textbf{Effect of language and code-switching.} \quad
A more pronounced pattern emerges across languages.
In both the raw ternary data and the binary preference subset, agreement on
Chinese (\textit{zh}) is noticeably higher than on English (\textit{en})
and especially on code-switched (\textit{mixed}) samples
(e.g., $74.4\%$ vs.\ $61.7\%$ and $62.5\%$ in
\text{SpeechJudge-Data (pref)}).
We attribute this gap to the annotator population:
our annotators are predominantly native Mandarin speakers; while English and
code-switched items are assigned only to annotators who pass a high English
proficiency bar, they are still L2 or bilingual listeners.
Similar phenomena---lower agreement for L2 or cross-lingual annotations
compared to L1 ones—have also been discussed in recent text RLHF studies~\citep{RLHF-anthoropic}.

In the current work, we therefore interpret SpeechJudge-Data as primarily
reflecting the preferences of {Chinese and Chinese--English bilingual}
listeners.
In future work, we plan to augment the corpus by recruiting
native English speakers for the English subset and
stronger bilingual/native speakers for code-switched data,
so as to improve agreement on these subsets and broaden the cultural and
linguistic coverage of the dataset.

\begin{table}[t]
\centering
\caption{Inter-annotator agreement (mean $\pm$ std, in \%) on SpeechJudge-Data and prior human preference datasets in the filed of RLHF.}
\vspace{-2mm}
\label{tab:inter-agreement}
\resizebox{\textwidth}{!}{
\begin{threeparttable} 
\begin{tabular}{l|c|cc|ccc|c}
    \toprule
    \multicolumn{1}{c|}{\multirow{2}{*}{\makecell[c]{ \textbf{Dataset}} }} &
    \multicolumn{1}{c|}{\multirow{2}{*}{\makecell[c]{ \textbf{Preference}} }} &
    \multicolumn{2}{c|}{\textbf{Regular}} &
    \multicolumn{3}{c|}{\textbf{Expressive}} &
    \multicolumn{1}{c}{\multirow{2}{*}{\makecell[c]{ \textbf{Total}} }} \\
    \cmidrule(lr){3-4} \cmidrule(lr){5-7}
    & & \textit{en} & \textit{zh} & \textit{en} & \textit{zh} & \textit{mixed} & \\
    \midrule
    \text{SpeechJudge-Data (raw)} & Ternary &
    54.9 $_{\scriptscriptstyle \pm \text{0.3\%}}$ &
    55.5 $_{\scriptscriptstyle \pm \text{0.2\%}}$ &
    50.2 $_{\scriptscriptstyle \pm \text{0.7\%}}$ &
    49.5 $_{\scriptscriptstyle \pm \text{0.4\%}}$ &
    36.3 $_{\scriptscriptstyle \pm \text{0.3\%}}$ &
    50.7 $_{\scriptscriptstyle \pm \text{0.1\%}}$ \\
    \midrule
    \text{SpeechJudge-Data (pref)} & \multirow{3}{*}{Binary} &
    61.7 $_{\scriptscriptstyle \pm \text{0.3\%}}$ &
    74.4 $_{\scriptscriptstyle \pm \text{0.2\%}}$ &
    59.9 $_{\scriptscriptstyle \pm \text{0.8\%}}$ &
    73.0 $_{\scriptscriptstyle \pm \text{0.6\%}}$ &
    62.5 $_{\scriptscriptstyle \pm \text{0.5\%}}$ &
    69.0 $_{\scriptscriptstyle \pm \text{0.2\%}}$ \\
    \text{ImageReward~\citep{imagereward}} &  & -- & -- & -- & -- & -- & 65.3 $_{\scriptscriptstyle \pm \text{5.6\%}}$ \\
    \text{InstructGPT~\citep{instructgpt}} &  & -- & -- & -- & -- & -- & 72.6 $_{\scriptscriptstyle \pm \text{1.5\%}}$ \\
    \bottomrule
\end{tabular}
\begin{tablenotes}
    \footnotesize{
        \item[*] {\text{SpeechJudge-Data (raw)} uses ternary labels (\text{A better} / \text{B better} / \text{Tie}); \text{SpeechJudge-Data (pref)} removes ties and uses binary labels (\text{A better} / \text{B better}). Results of SpeechJudge-Data are further broken down by style and language.}
    }
\end{tablenotes}
\vspace{-4mm}
\end{threeparttable}  
}
\end{table}

\subsection{Intelligibility Annotation Analysis}
\label{app:analysis-intelligibility}


We provide a detailed analysis of the relationship between the mostly common used objective intelligibility metric, Word Error Rate (WER), and the subjective human judgments of intelligibility. Our goal is to determine the extent to which WER can serve as a reliable proxy for human perception.

\begin{figure}[ht]
    \centering
    \includegraphics[width=1.0\linewidth]{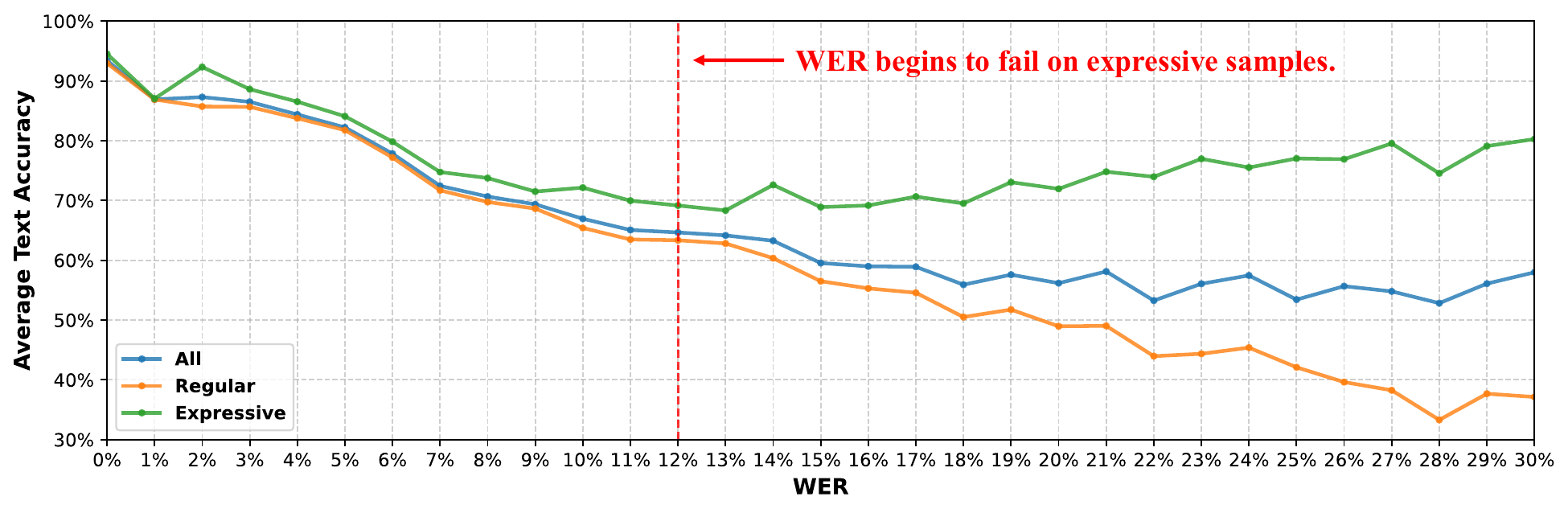}
    \caption{The relationship between the human-annotated text accuracy and WER.}
    \label{fig:wer_vs_error_binned}
\end{figure}

We use all the speech samples from SpeechJudge-Data (raw) for this analysis. We visualize the relationship between WER and the subjective text accuracy in Figure~\ref{fig:wer_vs_error_binned}.
For the regular speeches (the orange curve), we observe a consistent negative correlation: as the WER increases, its perceived text accuracy steadily declines. For the expressive speeches (the green curve), the similar trend holds for expressive speech when WER is under about 12\%. When WER is over the threshold, however, the correlation between WER and the subjective text accuracy weakens significantly. We think this divergence is sourced from that the greater stylistic variations in expressive speech pose a substantial challenge to the robustness of ASR systems compared to the regular samples.


\section{Details of Evaluation on the SpeechJudge-Eval Benchmark}
\label{app:details-of-evaluation-on-speechjudgeval}

During the evaluation on the SpeechJudge-Eval Benchmark of Table~\ref{tab:eval-benchmark}, we adopt the following protocol for each model:
\begin{itemize}[itemsep=1ex,leftmargin=3ex]
    \item \textbf{WER}~\citep{whisper,funasr}: We employ \texttt{Whisper-large-v3}\footnote{\href{https://huggingface.co/openai/whisper-large-v3}{https://huggingface.co/openai/whisper-large-v3}}~\citep{whisper} for English texts, and \texttt{Paraformer-zh}\footnote{\href{https://huggingface.co/funasr/paraformer-zh}{https://huggingface.co/funasr/paraformer-zh}}~\citep{paraformer, funasr} for Chinese and code-switching texts.
    \item \textbf{SIM}~\citep{wavlm}: We compute the cosine similarity between the WavLM TDNN\footnote{\href{https://github.com/microsoft/UniSpeech/tree/main/downstreams/speaker_verification}{https://github.com/microsoft/UniSpeech/tree/main/downstreams/speaker\_verification}}~\citep{wavlm} speaker embeddings of generated samples and the prompt samples.
    \item \textbf{FAD}~\citep{fad}: We use the officially released checkpoint, \texttt{VGGish}, to obtain the FADs of audios.
    \item \textbf{DNSMOS}~\citep{dnsmos}: We use the officially released script\footnote{\href{https://github.com/microsoft/DNS-Challenge/blob/master/DNSMOS/dnsmos\_local.py}{https://github.com/microsoft/DNS-Challenge/blob/master/DNSMOS/dnsmos\_local.py}} to calculate the DNSMOS of audios.
    \item \textbf{CE, CU, PC, and PQ}~\citep{meta-audiobox-aesthetics}: We use the officially released toolkit\footnote{\href{https://github.com/facebookresearch/audiobox-aesthetics}{https://github.com/facebookresearch/audiobox-aesthetics}} to get the predicted MOS for audios.
    \item \textbf{AASIST}~\citep{aasist}: It is a common baseline model for audio deepfake detection, employs a heterogeneity-aware approach to integrate spectral and temporal sub-graphs. We use a large-scale in-house corpus to train the model.
    \item \textbf{ADV}~\citep{audio-deepfake-verification}: It is a state-of-the-art (SOTA) deepfake detection model built upon the pre-trained \texttt{w2v-bert-2.0}\footnote{\href{https://huggingface.co/facebook/w2v-bert-2.0}{https://huggingface.co/facebook/w2v-bert-2.0}}. It utilizes a multi-task training approach involving deepfake source tracing to extract robust audio deepfake features. We use the same corpus of AASIST to train the model.
    \item \textbf{AudioLLMs}: We use the plain prompt of Table~\ref{tab:eval-protocol} to instruct AudioLLMs to pairwise score the naturalness of two audios. For the closed-source models, we use the official API released by Google\footnote{\href{https://ai.google.dev/gemini-api/docs/models}{https://ai.google.dev/gemini-api/docs/models}} for Gemini and OpenAI\footnote{\href{https://platform.openai.com/docs/models}{https://platform.openai.com/docs/models}} for GPT. We use the model variants \texttt{gemini-2.5-flash}, \texttt{gemini-2.5-pro}, \texttt{gpt-4o-mini-audio-preview-2024-12-17}, and \texttt{gpt-4o-audio-preview-2025-06-03} for Gemini-2.5-Flash, Gemini-2.5-Pro, GPT-4o mini Audio, and GPT-4o Audio.
\end{itemize}

\section{More Evaluation Results of Existing AudioLLMs}\label{app:more-results-of-audiollm-benchmark}

Using AudioLLM as a judge models, prompt engineering strategies are usually believed crucial for improving the performance~\citep{llm-as-a-judge,audiojudge}. Some common prompt engineering strategies include using the CoT prompts to activate the model's thinking and reasoning abilities~\citep{google-grm,deepseek-grm,audiojudge}, or employing few-shot evaluation formats~\citep{llm-as-a-judge,mimoaudio}.

\subsection{Chain-of-thought prompting for AudioLLM judges}\label{app:cot-prompt}

In this study, we investigate whether using the CoT from Table~\ref{tab:eval-protocol} helps AudioLLMs better judge speech naturalness. We display the results in Table~\ref{tab:eval-benchmark-cot}. Interestingly, we find that some closed-source AudioLLMs, such as Gemini-2.5-Flash, improve their performance on SpeechJudge-Eval through this thinking and reasoning process. However, this strategy often does not work for existing open-source AudioLLMs. For example, the results in Table~\ref{tab:eval-benchmark-cot} show that {Qwen2.5-Omni-7B} and Kimi-Audio-7B-Instruct, which is already the leading open-source models on SpeechJudge-Eval (Table~\ref{tab:eval-benchmark}), actually sees a decline in performance when using the CoT prompt.

\begin{table}[ht]
    \caption{Performance of AudioLLMs on SpeechJudge-Eval when using the CoT prompt of Table~\ref{tab:eval-protocol}.}
    \vspace{-2mm}
    \label{tab:eval-benchmark-cot}
    \centering
    \resizebox{0.6\textwidth}{!}{%
    \begin{threeparttable}
        \begin{tabular}{lccc}
            \toprule
             \multicolumn{1}{c}{\multirow{1}{*}{\textbf{Model}}}  & \multicolumn{1}{c}{\textbf{Regular}} & \multicolumn{1}{c}{\textbf{Expressive}} & \multirow{1}{*}{\textbf{Total}} \\
             \midrule
            {Qwen2.5-Omni-7B} & {62.0} & {59.7} & {60.6}  \\
            \multicolumn{1}{r}{{w/ CoT prompt}} & {54.4} & {50.6} & {52.9} \\
            \midrule
            {Kimi-Audio-7B-Instruct} & 65.5 & 68.0 & 67.0 \\
            \multicolumn{1}{r}{w/ CoT prompt} & 67.4 & 66.1 & 66.5 \\
            \midrule
            {Gemini-2.5-Flash} & 73.5 & 66.2 & 69.1 \\
            \multicolumn{1}{r}{w/ CoT prompt} & 75.0 & 67.5 & 70.5 \\
            \bottomrule
        \end{tabular}
    \end{threeparttable}    
    }   
    \end{table}

Based on our preliminary qualitative analysis, we believe the reason why the open-source AudioLLMs do not work well with the CoT prompt is that their foundational capabilities are relatively weak. These weaknesses include instruction-following (such as format-following), multiple-audio understanding, long-context processing, and reasoning abilities. This is also why, when we developed SpeechJudge-GRM, we did not directly apply RLVR on top of Qwen2.5-Omni-7B. Instead, we used an initial SFT stage as a cold start.

\subsection{Few-Shot Prompting for AudioLLM Judges}\label{app:few-shot-prompt}
Motivated by the common belief that in-context examples can improve LLM judging
ability~\citep{llm-as-a-judge,audiojudge}, we also investigate \textit{few-shot prompting} for Qwen2.5-Omni-7B on
SpeechJudge-Eval.

\begin{table}[ht]
    \caption{Performance of Qwen2.5-Omni-7B on SpeechJudge-Eval with different $k$-shot prompts.}
    \vspace{-2mm}
    \label{tab:eval-benchmark-few-shot}
    \centering
    \resizebox{0.65\textwidth}{!}{%
    \begin{threeparttable}
        \begin{tabular}{lccc}
            \toprule
            \multicolumn{1}{c}{\textbf{Model}}  &
            \multicolumn{1}{c}{\textbf{Regular}} &
            \multicolumn{1}{c}{\textbf{Expressive}} &
            \multicolumn{1}{c}{\textbf{Total}} \\
            \midrule
            {Qwen2.5-Omni-7B (0-shot)} & 62.0 & 59.7 & 60.6 \\
            \midrule
            \multicolumn{1}{r}{w/ 2-shot}  & 50.9 & 46.6 & 48.2 \\
            \multicolumn{1}{r}{w/ 4-shot}  & 46.1 & 52.0 & 49.6 \\
            \multicolumn{1}{r}{w/ 6-shot}  & 50.3 & 53.0 & 51.9 \\
            \multicolumn{1}{r}{w/ 8-shot}  & 51.0 & 54.8 & 53.3 \\
            \multicolumn{1}{r}{w/ 16-shot} & 48.6 & 53.0 & 51.3 \\
            \bottomrule
        \end{tabular}
    \end{threeparttable}
    }
\end{table}

We start from the \textit{plain} zero-shot prompt in Table~\ref{tab:eval-protocol},
which asks the model to decide which of two audios is more natural.
For the $k$-shot setting, we prepend $k$ preference exemplars to this prompt.
Each exemplar consists of:
(i) a target text,
(ii) an audio pair associated with this text, and
(iii) the corresponding human naturalness label (which of the two audios is
preferred).
The model is then queried on a new SpeechJudge-Eval pair with the same
instruction and output format.
We evaluate Qwen2.5-Omni-7B with $k\in\{2,4,6,8,16\}$; results are reported in
Table~\ref{tab:eval-benchmark-few-shot}.

Contrary to the usual expectation that few-shot prompting should help,
we observe that none of the $k$-shot configurations improves over the
zero-shot baseline.
On the contrary, the overall accuracy drops from $60.6\%$ in the 0-shot
setting to $48$--$53\%$ with few-shot prompts.
The degradation is particularly pronounced on regular speech, while expressive
speech shows small, inconsistent fluctuations.

These findings are consistent with our observations in
Appendix~\ref{app:cot-prompt} on chain-of-thought prompting:
current open-source AudioLLMs such as Qwen2.5-Omni-7B still have limited instruction-following, multi-audio
understanding, and long-context handling capabilities.
Although few-shot prompts provide more information in principle, the model
struggles to reliably associate multiple text-audio pairs with their labels in the context. As a result, increasing $k$ mainly adds complexity to the input without
yielding better judgments.
This further motivates our choice to move beyond pure prompt engineering,
and instead train dedicated reward models (BTRM / GRM) on human preference
data for robust naturalness evaluation.

\section{Training Details of SpeechJudge-GRM}\label{app:expt-setup}
\textbf{SFT Stage}\quad
We use Gemini-2.5-Flash~\citep{gemini2.5} to generate the CoT data for SpeechJudge-Data (train). For the total 42K samples, Gemini-2.5-Flash's judgments agree with human feedback on 25K samples, while they disagree on 17K samples. During the SFT stage, we fine-tune Qwen2.5-Omni-7B (Thinker)~\citep{qwen2.5-omni} on the 25K CoT data using LoRA~\citep{lora} with a rank of 128. We use Adam~\citep{adam,adamw} as the optimizer and set the learning rate to 5e-5. The maximum number of tokens per batch is 4000. We select the best checkpoint on SpeechJudge-Data (dev) as the SFT model, SpeechJudge-GRM (SFT).

\textbf{RL Stage}\quad
We use the 17K samples (as described above) to conduct DAPO~\citep{dapo}, which is an enhanced variant of GRPO~\citep{grpo}. We utilize the \texttt{ms-swift}\footnote{\href{https://github.com/modelscope/ms-swift}{https://github.com/modelscope/ms-swift}} toolkit to launch the training process. We initialize the policy model with the SFT model and use LoRA training with a rank of 64. The number of rollouts for each prompt is set to 8, and the batch size is 32. The learning rate is 5e-6. We select the best checkpoint on SpeechJudge-Data (dev) as the final SpeechJudge-GRM model, i.e., SpeechJudge-GRM (SFT+RL).

\section{More Evaluation Results for SpeechJudge-GRM}

\subsection{Performance under Different Data Distributions}

\begin{table}[t]
    \caption{Accuracy of different models on {SpeechJudge-Eval (regular)} across language settings.}
    \vspace{-2mm}
    \label{tab:eval-languages-regular}
    \centering
    \resizebox{0.75\textwidth}{!}{%
    \begin{threeparttable}
        \begin{tabular}{l|cc|cc|c}
            \toprule
            \multicolumn{1}{c|}{\multirow{2}{*}{\makecell[c]{ \textbf{Model}} }} &
            \multicolumn{5}{c}{\textbf{Regular}} \\
            \cmidrule(lr){2-6}
            & \textit{en2en} & \textit{zh2en} & \textit{zh2zh} & \textit{en2zh} & \textbf{Avg} \\
            \midrule
            Qwen2.5-Omni-7B & 48.1 & 58.5 & 75.7 & 66.0 & 61.0 \\
            Gemini-2.5-Flash & 59.4 & 62.8 & 81.6 & 87.6 & 72.8 \\
            \midrule
            SpeechJudge-BTRM & 66.0 & 71.3 & 86.4 & 86.6 & 77.5 \\
            SpeechJudge-GRM (SFT) & 67.0 & 74.5 & 84.5 & 85.6 & 77.8 \\
            \multicolumn{1}{r|}{w/ Voting@10} & 65.1 & 75.5 & 83.5 & 85.6 & 77.3 \\
            SpeechJudge-GRM (SFT+RL) & 69.8 & 77.7 & 85.4 & 83.5 & 79.0 \\
            \multicolumn{1}{r|}{w/ Voting@10} & 75.5 & 79.8 & 80.6 & 86.6 & 80.5 \\
            \bottomrule
        \end{tabular}
    \end{threeparttable}
    }
\end{table}

\begin{table}[t]
    \caption{Accuracy of different models on {SpeechJudge-Eval (expressive)} across language settings.}
    \vspace{-2mm}
    \label{tab:eval-languages-expressive}
    \centering
    \resizebox{0.95\textwidth}{!}{%
    \begin{threeparttable}
        \begin{tabular}{l|cc|cc|cc|c}
            \toprule
            \multicolumn{1}{c|}{\multirow{2}{*}{\makecell[c]{\textbf{Model}}}} &
            \multicolumn{7}{c}{\textbf{Expressive}} \\
            \cmidrule(lr){2-8}
            & \textit{en2en} & \textit{zh2en} & \textit{zh2zh} & \textit{en2zh} & \textit{en2mixed} & \textit{zh2mixed} & \textbf{Avg} \\
            \midrule
            Qwen2.5-Omni-7B            & 43.6 & 51.7 & 61.1 & 70.9 & 64.5 & 69.6 & 59.7 \\
            Gemini-2.5-Flash           & 53.6 & 68.3 & 73.3 & 76.4 & 64.5 & 67.1 & 66.2 \\
            \midrule
            SpeechJudge-BTRM           & 62.9 & 56.7 & 72.2 & 85.5 & 68.6 & 67.1 & 69.5 \\
            SpeechJudge-GRM (SFT)      & 61.4 & 66.7 & 89.1 & 77.8 & 74.4 & 73.4 & 73.7 \\
            \multicolumn{1}{r|}{w/ Voting@10} 
                                       & 69.3 & 75.0 & 88.9 & 90.9 & 71.1 & 74.7 & 77.8 \\
            SpeechJudge-GRM (SFT+RL)   & 71.4 & 65.0 & 81.1 & 86.4 & 70.2 & 81.0 & 76.0 \\
            \multicolumn{1}{r|}{w/ Voting@10} 
                                       & 75.0 & 66.7 & 82.2 & 89.1 & 72.7 & 84.8 & 78.7 \\
            \bottomrule
        \end{tabular}
    \end{threeparttable}
    }
\end{table}

In Table~\ref{tab:eval-benchmark} of the main paper, we reported accuracies on
{SpeechJudge-Eval} aggregated over all languages and styles.
Here we provide a more fine-grained view of how different judges behave under
different data distributions, and we additionally evaluate on a completely
out-of-distribution (OOD) test set involving real human speech versus
commercial TTS clones.

\textbf{Languages on the regular subset.}\quad
Table~\ref{tab:eval-languages-regular} breaks down performance on the
\textit{regular} subset of {SpeechJudge-Eval} by language setting:
\textit{en2en}, \textit{zh2en}, \textit{zh2zh}, and \textit{en2zh}.
Across all models we observe that pairs involving Chinese
(\textit{zh2zh} and \textit{en2zh}) are consistently easier than purely
English pairs (\textit{en2en}).
For example, SpeechJudge-GRM (SFT+RL) reaches $85.4$/$83.5\%$ accuracy on
\textit{zh2zh}/\textit{en2zh}, but only $69.8\%$ on \textit{en2en}, and the
same trend holds for BTRM and Gemini-2.5-Flash.
We believe several factors contribute to this gap.
On the data side, as shown in Appendix~\ref{app:inter-annotator}, Chinese
(\textit{zh}) subsets exhibit higher inter-annotator agreement than English
and mixed subsets, so supervision for English-like conditions is more varied and
harder to fit.
On the modeling side, current TTS systems may also produce relatively
high-quality English outputs compared to Chinese, making the naturalness differences between two
English samples more subtle and thus more difficult to judge reliably.
Even under these challenges, SpeechJudge-GRM (SFT+RL) with Voting@10 still
achieves the strongest average accuracy ($80.5\%$) among all open judges.

\textbf{Languages on the expressive subset.}\quad
Table~\ref{tab:eval-languages-expressive} shows the same breakdown for the
\textit{expressive} subset.
As expected, expressive speech is generally harder: all models are a few
points lower than on their regular counterparts.
The language pattern persists: Chinese-involving settings
(\textit{zh2zh}, \textit{zh2mixed}) tend to be easier than \textit{en2en}.
For instance, SpeechJudge-GRM (SFT+RL) with Voting@10 attains $82.2\%$ on
\textit{zh2zh} and $84.8\%$ on \textit{zh2mixed}, compared to $75.0\%$ on
\textit{en2en}.
This is consistent with the inter-annotator statistics in
Appendix~\ref{app:inter-annotator}, where expressive and code-switched
English subsets show lower human agreement, and also reflects the increased
linguistic and prosodic complexity of expressive and code-switched speech.
Overall, expressive data are slightly more challenging than regular data, but
the relative ranking of judges is stable and GRM maintains a clear advantage
over BTRM and the AudioLLM baselines across all language settings.

\begin{table}[t]
    \caption{Accuracy of different models on {SpeechJudge-Eval} across prompt styles.}
    \vspace{-2mm}
    \label{tab:eval-styles}
    \centering
    \resizebox{0.9\textwidth}{!}{%
    \begin{threeparttable}
        \begin{tabular}{l|ccccc}
            \toprule
            \multicolumn{1}{c|}{\textbf{Model}} &
            \textbf{Regular} &
            \textbf{Emotional} &
            \textbf{Accented} &
            \textbf{Whisper} &
            \textbf{Game} \\
            \midrule
            Qwen2.5-Omni-7B      & 61.0 & 56.7 & 64.4 & 57.1 & 61.3 \\
            Gemini-2.5-Flash     & 72.8 & 63.7 & 66.7 & 74.6 & 66.0 \\
            \midrule
            SpeechJudge-BTRM     & 77.5 & 69.8 & 71.3 & 76.2 & 66.8 \\
            SpeechJudge-GRM (SFT) & 77.8 & 75.3 & 80.5 & 71.4 & 70.2 \\
            \multicolumn{1}{r|}{w/ Voting@10} & 77.3 & 76.7 & 80.5 & 74.6 & 78.7 \\
            SpeechJudge-GRM (SFT+RL) & 79.0 & 74.4 & 81.6 & 79.4 & 74.5 \\
            \multicolumn{1}{r|}{w/ Voting@10} & 80.5 & 78.1 & 85.1 & 82.5 & 75.7 \\
            \bottomrule
        \end{tabular}
    \end{threeparttable}
    }
\end{table}

\textbf{Different prompt styles.}\quad
Table~\ref{tab:eval-styles} compares performance across prompt styles:
regular, emotional, accented, whisper, and game-character speech.
For SpeechJudge-BTRM we see that regular prompts are the easiest
($77.5\%$), while emotional and game styles are notably harder ($69.8\%$ and
$66.8\%$).
In contrast, SpeechJudge-GRM benefits more from the expressive settings:
with SFT+RL and Voting@10, GRM achieves $80.5\%$ on regular, but rises to
$85.1\%$ on accented and $82.5\%$ on whisper prompts, and narrows the gap on
emotional and game prompts (from about $10$ points for BTRM to only a few
points).
We believe this reflects the advantage of the generative reward model and its
CoT-based training: by explicitly reasoning about artifacts, prosody, and
style, GRM is better able to handle challenging conditions such as accented
speech and whisper, rather than overfitting to the most frequent regular style.

\textbf{OOD evaluation on real speech vs.\ commercial TTS clones.}\quad
Finally, Table~\ref{tab:eval-ood-real} presents a new, fully
out-of-distribution evaluation designed to stress-test generalization to
\textit{real} speech and unseen TTS architectures.
We select two native English voice actors, each recording 250 utterances
(500 in total), and use a commercial SeedTTS voice-cloning API\footnote{\href{https://www.volcengine.com/product/voicecloning}{https://www.volcengine.com/product/voicecloning}} to synthesize
clones for each utterance.
SeedTTS is a state-of-the-art proprietary system whose output quality is
typically higher than that of the open-source TTS models used to construct
{SpeechJudge-Data}, so this benchmark effectively probes the gap
between very strong modern TTS and human recordings.
For every sentence, we form a pair (human recording vs.\ SeedTTS clone) and
treat the human recording as the ground-truth more natural sample.
Neither the human recordings nor the SeedTTS outputs are present in
{SpeechJudge-Data}, making this a challenging OOD test.
The results show three interesting trends:

\begin{table}[t]
    \caption{Accuracy of different models on the out-of-distribution \textbf{human vs.\ SeedTTS clone} evaluation set.
Each test pair consists of a human recording and a SeedTTS voice clone for the same sentence; the model must decide which of the two is more natural.
In all pairs, the human recording is treated as the ground-truth more natural sample.}
    \vspace{-2mm}
    \label{tab:eval-ood-real}
    \centering
    \resizebox{0.7\textwidth}{!}{%
    \begin{threeparttable}
        \begin{tabular}{l|cc|c}
            \toprule
            \multicolumn{1}{c|}{\textbf{Model}} &
            \textbf{Character1} &
            \textbf{Character2} &
            \textbf{Avg} \\
            \midrule
            \rowcolor{gray!20} 
            \multicolumn{4}{c}{\textit{\textbf{Deepfake Detectors}}} \\
            AASIST & 97.2 & 100 & 98.6 \\
            ADV & 99.6 & 100 & 99.8  \\
            \midrule
            \rowcolor{gray!20} 
            \multicolumn{4}{c}{\textit{\textbf{AudioLLMs}}} \\
            Qwen2.5-Omni-7B          & 48.0 & 44.8 & 46.4 \\
            Kimi-Audio-7B-Instruct & 85.2 & 85.6 & 85.4 \\
            Gemini-2.5-Flash         & 52.8 & 48.8  & 50.8      \\
            \midrule
            \rowcolor{gray!20} 
            \multicolumn{4}{c}{\textit{\textbf{Naturalness Reward Model}}} \\
            SpeechJudge-BTRM         & 55.6 & 45.2 & 50.4 \\
            SpeechJudge-GRM (SFT)    & 37.6 & 44.0 & 40.8 \\
            \multicolumn{1}{r|}{w/ Voting@10} & 36.0 & 41.4 & 38.7 \\
            SpeechJudge-GRM (SFT+RL) & 57.6 & 67.2 & 62.4 \\
            \multicolumn{1}{r|}{w/ Voting@10} & 59.8 & 67.5 & 63.7 \\
            \bottomrule
        \end{tabular}
    \end{threeparttable}
    }
\end{table}

\begin{itemize}[itemsep=0ex,leftmargin=3ex]
\item First, deepfake detectors (AASIST and ADV) achieve almost perfect accuracy
(about $99\%$) on this task, even though they perform at chance level on
{SpeechJudge-Eval} (Table~\ref{tab:eval-benchmark}), confirming that they
mainly learn to discriminate real vs.\ synthetic rather than judge naturalness
between two synthetic samples.
\item Second, among AudioLLMs, Kimi-Audio-7B-Instruct performs strongly
($85.4\%$ on average), possibly because its training includes tasks or signals
related to authenticity detection. In contrast, Gemini-2.5-Flash attains only $50.8\%$ accuracy,
roughly at chance level and similar to Qwen2.5-Omni-7B ($46.4\%$).
This indicates that Gemini’s strong performance on {SpeechJudge-Eval}
(69.1\% in Table~\ref{tab:eval-benchmark}) does not automatically transfer to
the human--vs--clone setting: it appears to be a good judge for
\textit{synthetic-vs-synthetic} naturalness comparisons, but it is not explicitly
biased toward humans when facing a high-quality commercial clone.
\item Third, for naturalness reward models, SpeechJudge-BTRM is close to random
guessing ($50.4\%$), suggesting that the classical Bradley-Terry training may
overfit more heavily to the specific synthetic generators in
{SpeechJudge-Data}.
Interestingly, SpeechJudge-GRM (SFT) alone performs even worse than BTRM,
which we hypothesize is due to SFT encouraging the model to over-memorize our
prepared CoT patterns and thus hurting OOD generalization~\citep{gem,sft-rl-mayi}.
Once we add the RLVR stage, however, SpeechJudge-GRM (SFT+RL) improves
substantially to $62.4\%$ (and $63.7\%$ with Voting@10), outperforming BTRM
as well as generic AudioLLMs such as Qwen2.5-Omni-7B and Gemini-2.5-Flash on
this benchmark.
Given that the SeedTTS clones are already very close to human quality, this
performance indicates that SpeechJudge-GRM has the potential to provide useful
feedback not only for open-source TTS models, but also for strong proprietary
systems that were never seen during training.
This suggests that the generative reward modeling paradigm is more robust to
distribution shift than classical BTRM and off-the-shelf AudioLLM judges, and
that RL on human preferences is crucial for recovering and enhancing
generalization beyond the synthetic training distribution.
\end{itemize}

\subsection{Quality Analysis of Chain-of-Thought Reasoning}\label{app:grm-cot-quality}

Beyond scalar accuracy, we further analyze the \textit{quality} of the
Chain-of-Thought (CoT) rationales produced by different judges.
Specifically, we compare Gemini-2.5-Flash (teacher), SpeechJudge-GRM (SFT),
and SpeechJudge-GRM (SFT+RL) along three aspects:
(i) logical consistency between reasoning and conclusion,
(ii) faithfulness and hallucination rate as judged by human experts, and
(iii) differences in reasoning style.

\textbf{Consistency between reasoning and conclusion.}\quad
We first ask whether a model’s CoT reasoning is logically compatible with its
final decision.
Using DeepSeek-V3~\citep{deepseek-v3} as a meta-judge, we prompt it with the instruction like: given the CoT analysis of A and
B (over prosody, pacing, articulation, and overall naturalness) and the final
scores assigned to A and B, decide whether the conclusion is consistent with
the reasoning. DeepSeek-V3 returns a binary label (consistent / not consistent) and a brief
justification. The specific instruction is as follows:

\begin{promptbox}

\textit{\textbf{System Prompt:} }\\
I am having a model decide whether A or B is better using a Chain-of-Thought (CoT) process. Now, I need you to help me determine whether the model’s CoT output—its reasoning process and its final conclusion (which is presented as “A: X points, B: Y points”)—is consistent.

Please note:
\begin{itemize}[itemsep=0ex,leftmargin=3ex]
    \item You only need to check whether, in the CoT reasoning, if the model analyzes that A is worse than B, then in the final scores A should also be lower than B. As long as this condition is met, you can regard it as consistent. Conversely, if the model reasons that A is better than B but assigns A a lower score than B in the final output, then it is not consistent.
    \item You only need to return a JSON string with two keys.
    \begin{itemize}[itemsep=0ex,leftmargin=3ex,label=$\circ$]
        \item The first key is "result" with a value of 0 or 1—0 means not consistent, 1 means consistent.
        \item The second key is "reason", where you briefly explain your reasoning in English.
    \end{itemize}
\end{itemize}

\medskip \textit{\textbf{User Prompt (Example):} }\\

\textbf{Output A:}
\begin{itemize}[itemsep=0ex,leftmargin=3ex]
    \item \textbf{Prosody and Intonation:} The intonation is somewhat flat and lacks natural variability, making it sound a bit robotic. There are some unnatural stress patterns, and the overall prosody doesn't convey the meaning with human-like nuance.
    \item \textbf{Pacing and Rhythm:} The pacing is a bit slow and deliberate, and there's a slight unnatural pause before "They will frequently...". The rhythm feels somewhat monotonous.
    \item \textbf{Articulation and Clarity:} While most words are clear, there is a very noticeable and significant mispronunciation of "will" as "woof" ("they woof frequently make a show"). This is a major error that severely impacts naturalness.
    \item \textbf{Overall Naturalness:} The "woof" mispronunciation is a critical flaw that immediately breaks the illusion of natural speech. Combined with the somewhat flat prosody, the output sounds distinctly synthetic and flawed.
\end{itemize}

\textbf{Output B:}
\begin{itemize}[itemsep=0ex,leftmargin=3ex]
    \item \textbf{Prosody and Intonation:} The prosody is much more natural and varied, with appropriate pitch contours and stress patterns that align well with the meaning of the text. It sounds like a human speaker.
    \item \textbf{Pacing and Rhythm:} The pacing is excellent, flowing smoothly and naturally without any awkward pauses or rushed segments. The rhythm is consistent and pleasant to listen to.
    \item \textbf{Articulation and Clarity:} All words are articulated clearly and precisely. There are no noticeable mispronunciations, sibilance issues, or other artifacts.
    \item \textbf{Overall Naturalness:} This output sounds highly natural and human-like. The combination of good prosody, smooth pacing, and clear articulation results in a very convincing and pleasant listening experience.
\end{itemize}

\textbf{Conclusion:} Output A: 4, Output B: 8.5

\medskip \textit{\textbf{System Output (Example):} }\\
\{"result": 1, "reason": "In the reasoning, Output A is judged clearly worse than Output B (robotic prosody, major mispronunciation) while Output B is described as highly natural and human-like. The final scores reflect this ordering with A (4) $<$ B (8.5), so the conclusion is consistent with the analysis."\}

\end{promptbox}

Table~\ref{tab:cot-consistency} summarizes the results on
CoT outputs for SpeechJudge-Eval across the three models.
All three models exhibit very high internal consistency:
Gemini-2.5-Flash, SpeechJudge-GRM (SFT), and SpeechJudge-GRM (SFT+RL) achieve
$97.9\%$, $97.6\%$, and $98.2\%$ consistency, respectively.
This indicates that, at least at the coarse level captured by this automatic
check, the CoT analyses are not arbitrary narratives but align well with the
final naturalness preference.

\begin{table}[t]
    \caption{Consistency between Chain-of-Thought (CoT) reasoning and final
conclusion for different models on {SpeechJudge-Eval}, as judged by
DeepSeek-V3. A prediction is counted as consistent when the preference implied
by the CoT analysis matches the model’s final decision.}
    \vspace{-2mm}
    \label{tab:cot-consistency}
    \centering
    \resizebox{0.5\textwidth}{!}{%
    \begin{threeparttable}
        \begin{tabular}{l|c}
            \toprule
            \multicolumn{1}{c|}{\textbf{Model}} &
            \makecell[c]{\textbf{Consistency}} \\
            \midrule
            Gemini-2.5-Flash & 97.9\% \\
            SpeechJudge-GRM (SFT) & 97.6\% \\
            SpeechJudge-GRM (SFT + RL) & 98.2\% \\
            \bottomrule
        \end{tabular}
    \end{threeparttable}
    }
\end{table}

\begin{table}[t]
    \caption{Human-rated quality of Chain-of-Thought (CoT) explanations for
different models on {SpeechJudge-Eval}. Scores (1--3, $\uparrow$) are assigned by
expert raters on three dimensions—Prosody \& Intonation, Pacing \& Rhythm, and
Articulation \& Clarity—matching the dimensions specified in our CoT prompt (Table~\ref{tab:eval-protocol}).
We report the mean and its 95\% confidence interval for each dimension.}
    \vspace{-2mm}
    \label{tab:cot-human-quality}
    \centering
    \resizebox{0.85\textwidth}{!}{%
    \begin{threeparttable}
        \begin{tabular}{l|ccc|c}
            \toprule
            \multicolumn{1}{c|}{\textbf{Model}} &
            \makecell[c]{\textbf{Prosody}\\ \textbf{\& Intonation}} &
            \makecell[c]{\textbf{Pacing}\\ \textbf{\& Rhythm}} &
            \makecell[c]{\textbf{Articulation}\\ \textbf{\& Clarity}} &
            \textbf{Avg} \\
            \midrule
            Gemini-2.5-Flash
            & 1.90 $_{\scriptscriptstyle \pm \text{0.38}}$
            & 2.00 $_{\scriptscriptstyle \pm \text{0.44}}$
            & 2.10 $_{\scriptscriptstyle \pm \text{0.43}}$
            & 2.00 $_{\scriptscriptstyle \pm \text{0.34}}$ \\
            
            SpeechJudge-GRM (SFT)
            & 2.00 $_{\scriptscriptstyle \pm \text{0.39}}$
            & 2.15 $_{\scriptscriptstyle \pm \text{0.29}}$
            & 1.95 $_{\scriptscriptstyle \pm \text{0.42}}$
            & 2.03 $_{\scriptscriptstyle \pm \text{0.33}}$ \\
            
            SpeechJudge-GRM (SFT+RL)
            & 2.10 $_{\scriptscriptstyle \pm \text{0.20}}$
            & 2.00 $_{\scriptscriptstyle \pm \text{0.21}}$
            & 2.40 $_{\scriptscriptstyle \pm \text{0.32}}$
            & 2.17 $_{\scriptscriptstyle \pm \text{0.12}}$ \\
            \bottomrule
        \end{tabular}
    \end{threeparttable}
    }
\end{table}

\textbf{Human evaluation of CoT faithfulness.}\quad
Logical consistency does not guarantee that the reasoning is {correct} or
grounded in the audio.
To measure faithfulness and hallucination, we conduct a human evaluation with
experienced speech researchers (the background of these subjects are detailed in Appendix~\ref{app:sub-eval-details}).
For each model and each sampled {SpeechJudge-Eval} pair, the experts
examine the CoT and assign a 1--3 score on three dimensions—(1) Prosody and Intonation, (2) Pacing and Rhythm, and (3) Articulation and Clarity—which match the dimensions specified in our CoT prompt (Table~\ref{tab:eval-protocol}) and used by the models in their explanations.
A score of 3 means ``highly sensible (e.g., the CoT cites concrete audio details
and the analysis is correct)'', 2 means ``partially sensible (e.g., the overall
good/bad tendency is right but details are coarse or partially off)'', and 1
means ``not sensible / mostly hallucinatory''.

The results are reported in Table~\ref{tab:cot-human-quality}.
On average, all three models obtain scores around 2.0, indicating that their
CoT rationales are generally meaningful rather than dominated by hallucination.
Importantly, SpeechJudge-GRM does not lose CoT quality compared to the
Gemini teacher even though it is initialized from Gemini-generated rationales:
SpeechJudge-GRM (SFT) slightly improves the average score to $2.03\pm0.34$,
and after RL, SpeechJudge-GRM (SFT+RL) further increases it to
$2.17\pm0.12$.
The largest gain appears in the ``Articulation and Clarity'' dimension
($2.40$ vs.\ $2.10$ for Gemini), suggesting that RLVR encourages the model to
focus more accurately on pronunciation errors and intelligibility-related
artifacts when explaining its decisions.
Overall, the human study suggests that the RL stage not only improves
preference alignment but also mildly enhances the faithfulness of the CoT
reasoning.

\textbf{Differences in reasoning style.}\quad
Finally, we examine whether the three models share the same reasoning style or
develop distinct emphases.
We randomly sample 20 {SpeechJudge-Eval} cases on which all three models predict
the \textit{correct} preference label, and collect their CoT outputs.
We then submit these triplets of CoTs (anonymized as ``Model~1/2/3'') to three
strong text LLMs—GPT~5.1\footnote{\href{https://chatgpt.com/}{https://chatgpt.com/}}, Gemini~3~Pro\footnote{\href{https://aistudio.google.com/}{https://aistudio.google.com/}}, and DeepSeek-V3\footnote{\href{https://chat.deepseek.com/}{https://chat.deepseek.com/}}, asking them to compare the similarities
and systematic differences among the models. The specific instruction is as follows:

\begin{promptbox}

\textit{\textbf{System Prompt:} }\\
I would like you to analyze the reasoning processes in the Chain-of-Thought (CoT) outputs of the following three models (referred to as Model 1, Model 2, and Model 3, respectively). What are the similarities and significant pattern differences among them? I will provide you with 20 cases, and under each case, the outputs of Model 1, Model 2, and Model 3 will be given.

\medskip \textit{\textbf{User Prompt (Example):} }\\

******************** Case: 1/20 ********************
\vspace{2mm}

[Text] The rainbow serves as a metaphor for life's diversity - just as varying droplet sizes create different spectra, our unique experiences shape the breadth of our existence. The wider the band, the richer the experience.

[Model 1] ...

[Model 2] ...

[Model 3] ...

\vspace{2mm}
******************** Case: 2/20 ********************
\vspace{2mm}

[Text] They have committed to sending us the proposals to address our concerns by the end of January.

[Model 1] ...

[Model 2] ...

[Model 3] ...

...

\end{promptbox}

The three analyzers broadly agree on the following qualitative picture:
all models follow a similar structural template
(prosody $\rightarrow$ pacing $\rightarrow$ articulation $\rightarrow$
overall naturalness, followed by a 1--10 score), but they differ in what they
emphasize:

\begin{itemize}[itemsep=0ex,leftmargin=3ex]
\item \textbf{Gemini-2.5-Flash:}
  Described as the most sophisticated at linking prosody to {semantic}
  meaning and discourse structure, often explaining {why} an intonation
  pattern distorts the intended message.
\item \textbf{SpeechJudge-GRM (SFT):}
  Viewed as emotionally and narratively focused, with stable formatting and
  slightly more generous tone; it emphasizes human-likeness and expressiveness
  but is somewhat less detailed on low-level signal artifacts.
\item \textbf{SpeechJudge-GRM (SFT+RL):}
  Characterized as more critical and technically oriented: it pays more
  attention to mispronunciations, noise, and clarity, sometimes with blunter
  wording. Some analyzers note that its formatting is slightly less uniform
  than SpeechJudge-GRM (SFT), but its focus on error severity and technical correctness is
  stronger.
\end{itemize}

Taken together, these analyses suggest that SpeechJudge-GRM does not simply
imitate the teacher’s explanation style.
Instead, SFT initializes a shared analytical framework, while the RLVR stage
shifts the model toward rationales that are more tightly coupled to human
preferences and to concrete acoustic evidence
(especially articulation and clarity), without sacrificing internal
consistency.
We will release the full set of CoT examples and meta-analyses in our
open-source release to facilitate further research on CoT quality and
reasoning in audio reward models.

\subsection{Error Analysis of SpeechJudge-GRM}
\label{app:grm-error}

Although SpeechJudge-GRM performs well on {SpeechJudge-Eval}, it still
disagrees with the human on a subset of it.
We manually inspected these errors and found several recurring patterns.

\textbf{Over-weighting cleanness vs.\ liveliness.}\quad
In many errors the human-preferred sample contains mild background noise but clearly more human-like prosody and articulation, while the alternative
is cleaner yet more robotic or over-smoothed.
Annotators consistently favor the livelier sample, whereas GRM sometimes
chooses the cleaner one, indicating that it can over-emphasize acoustic
cleanness when the trade-off is subtle.

\textbf{Prosody-articulation trade-offs.}\quad
Another frequent pattern is a trade-off between expressive prosody and
technical correctness.
Humans often prefer speech with natural rhythm and intonation despite minor
pronunciation issues, while GRM occasionally favors the perfectly articulated
but flatter reading.
These cases reveal that the relative weighting between prosody and
articulation is still imperfectly captured.

\textbf{Extreme expressive styles.}\quad
Errors also concentrate in highly expressive styles.
For emotional speech with very high F0 or strong emphasis, humans interpret
the exaggerated prosody as appropriate for the style, but GRM sometimes
penalizes it as ``unnatural'' and prefers a neutral reading.
For whispers, the lack of voicing makes prosody judgments difficult; GRM
occasionally fails to distinguish the more fluent whisper when both samples
sound degraded.

\textbf{Very small preference gaps.}
A small number of mistakes arise when both clips are high-quality and differ
only in subtle cues (micro-pauses, breathing, slight emphasis shifts).
In these cases GRM’s predictions are effectively close to random, which is
unsurprising given the weak supervision signal.

In summary, SpeechJudge-GRM’s errors concentrate on nuanced trade-offs
(clean vs.\ lively, prosody vs.\ articulation) and on challenging expressive
styles, suggesting future work on modeling recording conditions, style-aware
priors, and more targeted training examples.

\section{High-Quality Sample Selection and Post-Training based on SpeechJudge-GRM}

\subsection{Details of Subjective Evaluation}\label{app:sub-eval-details}

During the construction of SpeechJudge-Data, we hired human labelers from a data crowdsourcing company. To verify the effectiveness of our training for them and to ensure the high quality of both the dataset and the resulting SpeechJudge-GRM, the human subjects for the final sample selection and TTS post-training experiments (Section \ref{sec:sample-selection} and \ref{sec:post-training-for-tts}) were all experienced speech generation researchers. All these researchers had extensive audio backgrounds, with a minimum of two years of experience in speech synthesis.

We randomly selected the subjective evaluation samples from both SeedTTS-Eval~\citep{seedtts}\footnote{\href{https://github.com/BytedanceSpeech/seed-tts-eval}{https://github.com/BytedanceSpeech/seed-tts-eval}} and Amphion-TTS-Eval~\citep{intp,vevo2}\footnote{\href{https://huggingface.co/datasets/amphion/Amphion-TTS-Eval}{https://huggingface.co/datasets/amphion/Amphion-TTS-Eval}}. The evaluation set for each system in Figure~\ref{fig:post-training-tts-results} consists of 70 samples, while the set for each system in Figure~\ref{fig:best-of-n-by-speechjudge-grm} contains 100 samples. Each audio sample in these evaluations received at least three independent ratings. These subjective evaluation results show that the annotation quality of SpeechJudge-Data largely aligns with the judgments of professional researchers.

\subsection{Objective Results}\label{app:post-training-tts}

\begin{table*}[t]
\caption{Post-training of Qwen2.5-0.5B-TTS based on SpeechJudge.}
\vspace{-3mm}
\label{tab:grm-post-trained-tts-details}
\begin{center}
\begin{threeparttable}
\resizebox{\textwidth}{!}{
\begin{tabular}{l||rr|rr|rr|rr||rr||rr}
\toprule
\multirow{2}{*}{\textbf{Model}} &
\multicolumn{2}{c|}{\textbf{Regular}} &
\multicolumn{2}{c|}{\textbf{Articulatory}} &
\multicolumn{2}{c|}{\textbf{Code-switching}} &
\multicolumn{2}{c||}{\textbf{Cross-lingual}} &
\multicolumn{2}{c||}{\textbf{Expressive}} &
\multicolumn{2}{c}{\textbf{Avg}} \\
\cmidrule(lr){2-3} \cmidrule(lr){4-5} \cmidrule(lr){6-7} \cmidrule(lr){8-9} \cmidrule(lr){10-11} \cmidrule(lr){12-13}
& \textbf{WER} & \textbf{SIM} & \textbf{WER} & \textbf{SIM} & \textbf{WER} & \textbf{SIM} & \textbf{WER} & \textbf{SIM} & \textbf{WER} & \textbf{SIM} & \textbf{WER} & \textbf{SIM} \\
\midrule
\multicolumn{1}{l||}{\text{Qwen2.5-0.5B-TTS}} & 2.63 & 0.698 & 10.53 & 0.679 & 23.87 & 0.666 & 10.51 & 0.593 & 11.10 & 0.706 & 11.73 & 0.668 \\
\midrule
\multicolumn{1}{r||}{\text{w/ INTP}} & 2.06 & 0.697 & 8.62 & 0.694 & 18.37 & 0.663 & 7.12 & 0.588 & 9.80 & 0.708 & 9.19 & 0.670 \\
\multicolumn{1}{r||}{\text{w/ SpeechJudge-Data}} & 2.12 & 0.698 & 8.92 & 0.678 & 19.01 & 0.657 & 7.72 & 0.583 & 9.97 & 0.707 & 9.55 & 0.664 \\
\multicolumn{1}{r||}{\text{w/ SpeechJudge-GRM (offline)}} & 2.31 & 0.698 & 7.83 & 0.681 & 15.36 & 0.662 & 7.84 & 0.593 & 9.72 & 0.709 & 8.51 & 0.668 \\
\multicolumn{1}{r||}{\text{w/ SpeechJudge-GRM (online)}} & 2.35 & 0.696 & 8.45 & 0.674 & 15.87 & 0.653 & 7.82 & 0.580 & 9.79 & 0.702 & 8.85 & 0.661 \\
\bottomrule
\end{tabular}
}
\end{threeparttable}
\end{center}
\end{table*}

We present the objective results (WER and SIM) of the Qwen2.5-0.5B-TTS post-training in Table~\ref{tab:grm-post-trained-tts-details}. The results show that all four post-training methods significantly improve the WER. This trend is similar to the subjective intelligibility results shown in Figure~\ref{fig:post-train-tts-naturalness}.

Regarding the SIM metric, both w/ INTP and w/ SpeechJudge-GRM (offline) either match or slightly outperform the baseline model, while the other two methods show a slight decline. However, the objective SIM results appear to be in slight conflict with the subjective speaker similarity results in Figure~\ref{fig:post-train-tts-speaker-similarity}. For instance, in the subjective evaluation, w/ INTP actually shows a decrease in speaker similarity (Win: 24.30\%, Lose: 32.90\%).

Through follow-up interviews with the subjects who participated in our subjective evaluation, we gathered additional qualitative insights. Participants consistently reported that the synthesized samples, both before and after post-training, demonstrated excellent speaker similarity, closely matching the reference speaker's timbre and style. In most cases, participants found it challenging to distinguish any significant differences in similarity, leading them to prefer selecting ``Tie". For example, in Figure~\ref{fig:post-train-tts-speaker-similarity}, all four methods have the highest ``Tie" proportion, each exceeding 40\%. This demonstrates that post-training methods centered on naturalness (SpeechJudge-based) or intelligibility (INTP-based) are not yet fully aligned with speaker similarity, which requires further research into speaker similarity alignment.

\end{document}